\documentclass[12pt]{article}
\usepackage[utf8x]{inputenc}
\usepackage[T1]{fontenc}

\usepackage{hfstyle}

\usepackage{multirow}
\usepackage[all,cmtip]{xy}
\usepackage{amsmath, amssymb,bbm,color}

\newtheorem{prop}{Proposition}[section]
\newtheorem{cor}{Corollary}[section]

\newtheorem{thm}{Theorem}[section]

\newcommand{\mfrac}[2]{\genfrac{}{}{1.0pt}{}{#1}{#2}}

\usepackage[draft,firstpage]{draftmark}
\draftmarksetup{fontfamily=ptm,angle=0,scale=0.07,mark={\mbox{Updated version of the paper published in \emph{Journal of Cellular Automata} 7, pp. 455-488 (2013)}},color=red,xcoord=-78,ycoord=120}

\begin{document}
\title{Construction of local structure maps for cellular automata}
\author{Henryk Fuk\'s 
      \oneaddress{
         Department of Mathematics\\
            Brock University\\
         St. Catharines, Ontario  L2S 3A1, Canada\\
         \email{hfuks@brocku.ca}
       }
   }
\date{}
\Abstract{
The paper formalizes and extends the idea of local structure approximation for cellular
automata originally proposed by Gutowitz \emph{et. al.} \cite{gutowitz87a}.
We start with a review of the construction of a probability measure on the set of bi-infinite
strings over a finite alphabet of $N$ symbols. We then demonstrate that for a shift-invariant probability measure,
probabilities of all blocks of length up to $k$ can be expressed by $(N-1)N^{k-1}$ linearly independent block probabilities.
 Two  choices of these independent blocks are discussed in detail,
one in which we choose the longest possible blocks (``long form'') and one in which we choose
the shortest possible blocks (``short form''). We then proceed to review the method
which allows to approximate probabilities of blocks longer than $k$ by blocks of length $k$
or less. This approximation, known as Bayesian extension or Markov measure, is then used
to construct approximate orbits of shift-invariant probability measures under the action
of probabilistic or deterministic cellular automaton. We show that the aforementioned approximate orbit is 
completely determined by an $(N-1)N^{k-1}$-dimensional map.
When the short form of block probabilities is used, this map takes particularly simple
form, often revealing important features of  a particular cellular automaton.
}

\maketitle

\section{Introduction}
Cellular automata (CA) are often considered as maps in the space of Borel 
shift-invariant probability measures  equipped
with the weak$\star$ topology \cite{KurkaMaas2000,Kurka2005,Pivato2009,FormentiKurka2009}.
The central problem of the theory of cellular automata in this setting is to determine
properties of orbits of  given initial measures $\mu$ under the action of a given cellular automaton.
Since computing the orbit of a measure is in general very difficult, approximate methods have been considered.
The simplest of these methods is called the mean-field theory, and has its origins in statistical physics \cite{Wolfram83}.
The main idea behind the mean-field theory is to approximate the consecutive iterations of 
the initial measure by  Bernoulli measures. While this approximation is obviously very crude,
it is sometimes quite useful in applications.

In 1987,  H. A. Gutowitz, J. D. Victor, and B. W. Knight \cite{gutowitz87a} proposed a generalization
of the mean-field theory for cellular automata which, unlike mean-field theory,  takes (partially) into account 
correlations between sites. The basic idea of local structure theory is to consider
probabilities of blocks of length $k$ and to construct a map on these block probabilities, which, when iterated,
approximates probabilities of occurrence of the same  blocks in the actual orbit of a given  cellular automaton.
The construction was based on  the idea of  ``Bayesian extension'', introduced earlier by other  authors
in the context of lattice gases
\cite{Brascamp71,Fannes84}, and also known as a ``finite-block measure'' or as ``Markov process with memory''.

In the original paper, Gutowitz \emph{et. al.} made a compelling argument that ``the local structure theory appears to be a powerful
method for characterization and classification of cellular automata'' \cite{gutowitz87a}. After performing
extensive Monte-Carlo simulations and statistical analysis they concluded that the local structure 
``is an accurate model of several aspects of cellular automaton evolution. The dependence on initial conditions
and convergence properties are well modeled by the theory. It appears that, even for complex rules,
the stable invariant measures of a cellular automaton may be estimated to arbitrary resolution''~\cite{gutowitz87a}.

In the last 25 years, the local structure theory has been applied to study various aspects of 
dynamics of both deterministic and probabilistic cellular automata, including, for example,
such topics as classification of CA, phase transitions in probabilistic CA, CA models of traffic flow, 
asynchronous CA, and many others. In spite of this, there has been virtually no effort to study this theory
from a more formal point of view, in order to obtain rigorous results which could be confronted with
 Monte Carlo experiments and numerical simulations. This paper is intended to be a fist step toward
filling this gap. 

The paper is organized as follows. In the first section, we review the classic construction of measures
on ${\mathcal{A}}^\mathbb{Z}$, where ${\mathcal{A}}=\{0,1,\ldots,N-1\}$, 
using cylinder sets and the Hahn-Kolmogorov extension theorem. In the next section
we show that for a shift-invariant measure, measures of all cylinders sets of length up to $k$, 
which we call ``block probabilities'',  can be generated  by $(N-1)N^k$ linearly independent block probabilities.
 We describe two choices of these independent blocks probabilities, ``long form'' and ``short form''.
In section 4 we show how the knowledge of measures of cylinder sets of length up to $k$ can be used to
approximate the entire measure. This construction is sometimes known as the ``maximal entropy'' extension.
We present proof of the maximality of the entropy following the idea given in \cite{frechet15} and
adopted to our formalism.

The maximal entropy extension is then used to construct approximate orbit of a measure $\mu$
under the action of cellular automaton.  Points of this orbit are entirely determined
by  $(N-1)N^k$ block probabilities, thus it is possible to generate approximate orbits by iterating
$(N-1)N^k$-dimensional real maps, instead of much more complicated $N^k$ dimensional maps
proposed in \cite{gutowitz87a}.   We also show that, as $k$ increases, every point of the approximate
orbit weakly converges to the corresponding point of the exact orbit.  

Finally, we present some examples of local structure maps and their reduced form.

\section{Construction of a probability measure}
Let ${\mathcal{A}}=\{0,1,\ldots,N-1\}$ be called an \emph{alphabet}, or a \emph{symbol set}, 
and let $X={\mathcal{A}}^\mathbb{Z}$.
The Cantor metric on $X$ is defined as $d(\mathbf{x},\mathbf{y})=2^{-k}$, where $k=\mathrm{min} \{ |i|: \mathbf{x}_i
\neq  \mathbf{y}_i\}$. $X$ with the metric $d$ is a Cantor space, that is, compact, totally
disconnected and perfect metric space.
A finite sequence of elements of ${\mathcal{A}}$, $\mathbf{b}=b_1b_2\ldots, b_{n}$ will be called a \emph{block}  (or \emph{word})
 of length $n$.
Set of all blocks of elements of ${\mathcal{A}}$ of all possible lengths will be denoted by ${\mathcal{A}}^{\star}$.
\emph{Elementary cylinder set} generated by the block $\mathbf{b}=b_1b_2\ldots, b_{n}$ and anchored at $i$  is defined as
\begin{equation}
[\mathbf{b}]_i=\{ \mathbf{x}\in {\mathcal{A}}^\mathbb{Z}: \mathbf{x}_{[i,i+n)}=\mathbf{b} \},
\end{equation}
where we require that one of the indices $i,i+1,\ldots, i+n-1$ is equal to zero, or, equivalently,
that $-n+1 \leq i \leq 0$.
For a given elementary cylinder set $[\mathbf{b}]_i$, indices  $i,i+1,\ldots, i+n-1$ will be called
\emph{fixed}, while all other indices will be called \emph{free}. 
The requirement
$-n+1 \leq i \leq 0$, therefore, means that the origin is always fixed.\footnote{Other 
choices of elementary cylinder sets are possible, not requiring fixed origin -- see, for
example, \cite{FormentiKurka2009}. Our choice is motivated by the desire that the set of elementary cylinder sets is
closed under intersection.}
The collection (class) of all elementary cylinder sets of $X$ together with the
empty set and the whole space $X$ will be denoted by $\mathit{Cyl}(X)$. We will
use the convention that for $\mathbf{b}=\varnothing$, $[\mathbf{b}]_i=X$.

Let $[\mathbf{a}]_j$ and $[\mathbf{b}]_i$ be two elementary cylinder sets.
We will say that $p\in \mathbb{Z}$ is a \emph{matching} (\emph{mismatching}) index of these cylinder 
sets if for every $\mathbf{x}
\in [\mathbf{a}]_j$, $\mathbf{y} \in [\mathbf{b}]_i$ we have $x_p=y_p$  
($x_p \neq y_p$). An index which is either matching or mismatching will be called
\emph{overlapping} index. Note that since we require that the origin is fixed, any two
cylinder sets must have at least one overlapping index.
\begin{prop}
 The collection of all elementary cylinder sets together with the empty set and the whole space constitutes a semialgebra
over $X$.
\end{prop}
\emph{Proof:}
In order to show that elementary cylinder sets constitute a semialgebra we  need to prove (i) the closure under 
the intersection and (ii) that the set difference of two elementary cylinder sets is a finite union of elementary
cylinder sets.

For (i), let $[\mathbf{a}]_j$  and $[\mathbf{b}]_i$ be two elementary cylinder sets.
As such, they must have some overlapping indices. If all overlapping indices are matching,
then $[\mathbf{a}]_j \cap [\mathbf{b}]_i$
is just the elementary cylinder set generated by overlapped concatenation of $\mathbf{a}$
and $\mathbf{b}$. If among overlapping indices there is at least one
mismatching index, then $[\mathbf{a}]_j \cap [\mathbf{b}]_i$ is empty.

For (ii), let us observe that 
\begin{equation}
 X \setminus  [\mathbf{b}]_i =\bigcup_{j=0}^{n-1} \{\mathbf{x}\in X : x_{i+j} \neq b_{j+1}\}.
\end{equation}
Each of the sets $ \{\mathbf{x}\in X : x_{i+j} \neq b_{j+1}\}$ can be expressed as a union of elementary cylinder
sets, thus $X \setminus  [\mathbf{b}]_i$ is also a union of elementary cylinder
sets. Now, since
\begin{equation}
 [\mathbf{a}]_j \setminus [\mathbf{b}]_i=[\mathbf{a}]_j \cap \big(X \setminus  [\mathbf{b}]_i\big),
\end{equation}
and (i) holds, we obtain the desired result.  $\square$

We will now introduce the notion of a measure on the semi-algebra of cylinder sets.
Let $\mathcal{D}$ be a semialgebra. A map $\mu: {\mathcal{D}} \to [0,\infty]$ is called a \emph{measure} on $\mathcal{D}$ if 
it is countably additive and $\mu(\varnothing)=0$. By \emph{countable additivity} we mean that for any sequence
$\{A_i\}_{i=1}^{\infty}$ of pairwise disjoint sets belonging to  $\mathcal{D}$ such that $\bigcup_{i=1}^{\infty} A_i \in {\mathcal{D}}$,
\begin{equation}
\mu\left(\bigcup_{i=1}^{\infty}A_i\right) = \sum_{i=1}^{\infty} \mu(A_i).
\end{equation}
For measures on the semialgebra of cylinder sets, countable additivity is implied by finite additivity.
\begin{prop}\label{finaddmap}
 Any finitely additive map $\mu: \mathit{Cyl}(X) \to [0,\infty]$  for which  $\mu(\varnothing)=0$ 
is a measure on the semialgebra of elementary cylinder sets $\mathit{Cyl}(X)$.
\end{prop}
\emph{Proof:}
We start with a remark that in the Cantor topology elementary
cylinder sets are clopen, that is, both closed and open. 

Suppose now that the
map $\mu$ satisfies  $\mu(\varnothing)=0$ and is finitely additive, that is,
for any finite sequence
$\{A_i\}_{i=1}^{m}$ of pairwise disjoint sets belonging to  $\mathit{Cyl}(X)$ such that 
$\bigcup_{i=1}^{m} A_i \in {\mathit{Cyl}(X)}$,
\begin{equation}
\mu\left(\bigcup_{i=1}^{m}A_i\right) = \sum_{i=1}^{m} \mu(A_i).
\end{equation}
In order to show that $\mu$ is a measure on $\mathit{Cyl}(X)$, we need to show that it is
countably additive. Let $B$ be a cylinder set and let $\{A_i\}_{i=1}^{\infty}$ be a collection
of pairwise disjoint cylinder sets such that $\bigcup_{i=1}^{\infty}A_i =B$. Since $B$ is closed, it is
also compact. Sets $A_i$ are open, and form a cover of the compact set $B$. There must exist,
therefore, a finite subcover, that is, a finite number of sets $A_i$ covering $B$.  
Moreover, since $A_i$ are mutually disjoint, there must exist $m$ such that 
$A_i=\varnothing$ for all $i>m$, and therefore $B=\bigcup _{i=1}^{m}A_i$. Then by finite
additivity of $\mu$ and the assumption that $\mu(\varnothing)=0$ we obtain
\begin{equation}
 \mu(B)=\mu\left(\bigcup_{i=1}^{\infty}A_i\right)=
\mu\left(\bigcup_{i=1}^{m}A_i\right)=
\sum_{i=1}^{m}\mu(A_i)= \sum_{i=1}^{\infty}\mu(A_i),
\end{equation}
which means that $\mu$ is countably additive and thus is a measure on $\mathit{Cyl}(X)$, as required. $\square$

Although the above proposition allows us to introduce a measure on the semialgebra of elementary cylinder set,
this semialgebra is ``too small'' a class of subsets of $X$ to support the full machinery
of probability theory. For this we need a $\sigma$-algebra, that is, a class
of subsets of $X$ that is closed under the complement and countable unions of its members.  
Such $\sigma$-algebra can be defined as an ``extension'' of $\mathit{Cyl}(X)$. The smallest 
$\sigma$-algebra containing $\mathit{Cyl}(X)$ will be called \emph{$\sigma$-algebra generated
by $\mathit{Cyl}(X)$}. As it turns out, it is possible to extend a measure on semi-algebra to
the $\sigma$-algebra generated by it, as the following theorem attests.  
\begin{thm}[Hahn-Kolmogorov]\label{hktheorem}
 Let $\mu: {\mathcal{D}} \to [0,\infty]$ be a measure on semi-algebra ${\mathcal{D}}$ of subsets of a set $Y$. Then
$\mu$ can be extended to a measure on the $\sigma$-algebra generated by $\mathcal{D}$. 
\end{thm}

This classic result  has been first proved by M. Fréchet \cite{frechet15}, and later
by A. Kolmogorov \cite{kolmogorov1933grundbegriffe} and H. Hahn \cite{hahn33}.
One can find its contemporary proof in ref. \cite{Parthasarathy77}. The proof 
is based on construction of the outer measure $\mu^\star$ determined by $\mu$, and 
then applying Carathéodory's extension theorem. Since the proof bears little relevance
to our subsequent considerations, it will be omitted here.

One can also show that the extension is unique if $\mu$ satisfies additional conditions. 
Without discussing this issue in full generality, we will only state that for 
probabilistic measures, that is, measures satisfying $\mu(X)=1$, the extension is
always unique \cite{Parthasarathy77}. In all subsequent considerations,
we will assume that the measure is probabilistic, and the set of all probabilistic measures
on the $\sigma$-algebra generated by elementary cylinder sets of $X$ will be
denoted by $\mathfrak{M}(X)$.

The Hahn-Kolmogorov Theorem  coupled with Proposition \ref{finaddmap} results in the following corollary,
which summarizes our discussion.
\begin{cor}
 Any finitely additive map $\mu: \mathit{Cyl}(X) \to [0,1]$ satisfying  $\mu(\varnothing)=0$ 
and $\mu(X)=1$ extends uniquely to a measure on the $\sigma$-algebra generated by 
elementary cylinder sets of $X$.
\end{cor}

The last thing we need to do is to characterize finite additivity of maps on $\mathit{Cyl}(X)$
in somewhat simpler terms. Recall that  $\mu: \mathit{Cyl}(X) \to [0,1]$ is finitely additive 
if for every  $B\in \mathit{Cyl}(X)$ and pairwise disjoint  $A_i \in \mathit{Cyl}(X)$, $i=1,2,\ldots,m$ such that
 $B=\bigcup_{i=1}^m A_i$, we have $\mu(B)=\sum_{i=1}^m \mu(A_i)$.
If $B$ is a cylinder set, when could it be a union of a finite number of
other cylinder sets, pairwise disjoint? From the definition of the cylinder set,
it is clear that if $B$ is a finite union of $A_i$, then each $A_i$ must be longer
than $B$, and for each pair $(B, A_i)$ all fixed indices of $B$ must be matching.
For $B=[\mathbf{b}]_i$ this can happen in one of the following three situations:
\begin{align}
 [\mathbf{b}]_i&=\bigcup_{\mathbf{a}\in {\mathcal{A}}^k} [\mathbf{ba}]_i,\label{decomp1}\\
[\mathbf{b}]_i&=\bigcup_{\mathbf{a}\in {\mathcal{A}}^k}  [\mathbf{ab}]_{i-k},\label{decomp2}\\
[\mathbf{b}]_i&=\bigcup_{\mathbf{a}\in {\mathcal{A}}^k,\,\, \mathbf{c}\in {\mathcal{A}}^l} 
[\mathbf{bac}]_i.\label{decomp3}
\end{align}
This means that we attach to $\mathbf{b}$ a postfix word, a prefix word, or both, and take the union over 
all values of attached word(s). Note that all cylinder sets on the right hand side of each
of the above equations are pairwise disjoint. If we want to test the map for countable
additivity, it is thus sufficient to test it on cases described by equations (\ref{decomp1}--\ref{decomp3}).
\begin{prop}\label{propadditivitysimple}
 The map  $\mu: \mathit{Cyl}(X) \to [0,1]$ is countably additive if and only if
for all $[\mathbf{b}]_i \in \mathit{Cyl}(X) \setminus X$,
\begin{equation}\label{additivitysimple}
 \mu([\mathbf{b}]_i)=\sum_{a\in {\mathcal{A}}} \mu ([\mathbf{b}a]_i)=
\sum_{a\in {\mathcal{A}}} \mu ([a\mathbf{b}]_{i-1}).
\end{equation}
\end{prop}
\emph{Proof:} Suppose that the map is countably additive. Applying additivity condition to
eq. (\ref{decomp1}) and (\ref{decomp2}) with $\mathbf{a}=a$ yields the desired result.

Now suppose that the double equality (\ref{additivitysimple}) holds. Applying it recursively
$k$ times we obtain
\begin{equation}
 \mu([\mathbf{b}]_i)=\sum_{a_1\in {\mathcal{A}}}\ldots \sum_{a_k\in {\mathcal{A}}}\mu([\mathbf{b}a_1a_2\ldots a_k]_i)=
\sum_{\mathbf{a}\in {\mathcal{A}}^k} \mu([\mathbf{ba}]_i),
\end{equation}
which implies additivity of $\mu$ for the case covered by eq. (\ref{decomp1}). One can deal with
cases covered by eqs. (\ref{decomp2}) and (\ref{decomp3}) in a similar fashion. The map $\mu$ 
is thus countably additive on $\mathit{Cyl}(X)$. $\square$

Note that when $\mathbf{b}=\varnothing$, according to our convention, $[\mathbf{b}]_i=X$,
and eq. (\ref{additivitysimple}) reduces to
\begin{equation}
 \sum_{a \in {\mathcal{A}}} \mu([a]_i)=1,
\end{equation}
where we used the assumption that the measure is probabilistic, $\mu(X)=1$.

\section{Shift-invariant measure}
In the previous section we demonstrated that any map $\mu: \mathit{Cyl}(X) \to [0,1]$ satisfying
$\mu(\varnothing)=0$, $\mu(X)=1$ and conditions of eq. (\ref{additivitysimple})
extends uniquely to a measure on the $\sigma$-algebra generated by 
elementary cylinder sets of $X$. We will now impose another condition on the
map $\mu: \mathit{Cyl}(X) \to [0,1]$, namely translational invariance (also called shift-invariance),
by requiring that, for all $\mathbf{b} \in {\mathcal{A}}^{\star}$,  $\mu([\mathbf{b}]_i)$ is 
independent of $i$. To simplify notation, we then define $P: {\mathcal{A}}^{\star} \to [0,1]$ as
\begin{equation}
 P(\mathbf{b}):=\mu([\mathbf{b}]_i).
\end{equation}
Values $P(\mathbf{b})$ will be called \emph{block probabilities}.
Applying Proposition \ref{propadditivitysimple} and Hahn-Kolmogorov theorem to the case
of shift-invariant $\mu$ we obtain the following result.
\begin{thm}\label{extensionfromblocks}
 Let $P: {\mathcal{A}}^{\star} \to [0,1]$  satisfy
the conditions
\begin{align} 
P(\mathbf{b})&= \sum_{a \in {\mathcal{A}}} P(\mathbf{b}a) =\sum_{a \in \cal{G}} P(a\mathbf{b})
\,\,\,\,\,\,\,\forall {\mathbf{b} \in {\mathcal{A}}^{\star}} ,\label{cons1}\\
1&=\sum_{a \in \cal{G}} P(a).\label{cons2}
\end{align}
Then $P$ uniquely determines shift-invariant probability measure on the $\sigma$-algebra
generated by elementary cylinder sets of $X$.
\end{thm}
The set of shift-invariant probability measures on the $\sigma$-algebra
generated by elementary cylinder sets of $X$ will be denoted by $\mathfrak{M}_\sigma(X)$.
Conditions (\ref{cons1}) and (\ref{cons2}) are often called \emph{consistency conditions}.
It should be stressed, however, they they are essentially equivalent to measure additivity
conditions. Nevertheless, since the term ``consistency conditions'' is
prevalent in the literature, we will use it in the subsequent considerations.

Since $P$ uniquely determines the probability measure, we can use block probability values
to define shift-invariant probability measure. Obviously, because of
consistency conditions, block probabilities are not independent. 

We will define $\mathbf{P}^{(k)}$ to be the column vector of all 
probabilities of blocks of length $k$ arranged in lexical order. For
example, for ${\mathcal{A}}=\{0,1\}$, these are
\begin{align*}
 \mathbf{P}^{(1)}&=[P(0), P(1)]^T,\\
\mathbf{P}^{(2)}&=[P(00),P(01),P(10),P(11)]^T,\\
\mathbf{P}^{(3)}&=[P(000),P(001),P(010),P(011),P(100),P(101),P(110),P(111)]^T,\\
&\cdots .
\end{align*}
Using this notation, eq. (\ref{cons1}) can be written as 
\begin{equation}\label{matrixcons}
 \mathbf{P}^{(k-1)}=
\mathbf{R}^{(k)}\mathbf{P}^{(k)}=\mathbf{L}^{(k)}\mathbf{P}^{(k)},
\end{equation}
where $k>1$ and where
 $\mathbf{L}^{(k)}$ and $\mathbf{R}^{(k)}$ are binary matrices
with $N^{k-1}$ rows and $N^k$ columns. In order to describe structure of these matrices,
let us denote identity matrix $N^{k-1}\times N^{k-1}$ by $\mathbf{I}$, 
and let $\mathbf{J}_{m}$ be a  $N^{k-1}\times N^{k-1}$ matrix in which $m$-th
row consist of all 1's, and all other entries are 0. Then 
$\mathbf{L}^{(k)}$ and $\mathbf{R}^{(k)}$ can be written as
\begin{align}
 \mathbf{L}^{(k)}&=[\,\underbrace{\mathbf{I}\,\,\, 
\mathbf{I} \ldots \mathbf{I}}_{N}\,], \label{defL}\\
 \mathbf{R}^{(k)}&=[\mathbf{J}_{1} \mathbf{J}_{2} \ldots \mathbf{J}_{N}].\label{defR}
\end{align}
For example, for $N=3$, we have
\begin{align}
 \mathbf{P}^{(2)}&=[P(00), P(01), P(02), P(10), P(11), P(12), P(20), P(21), P(22)]^T,\\
\mathbf{P}^{(1)}&=[P(0), P(1), P(2)]^T,
\end{align}
and eq. (\ref{matrixcons}) for $k=2$ becomes
\begin{equation}
\mathbf{P}^{(1)} =
\left[ \begin {array}{ccc|ccc|ccc} 1&1&1&0&0&0&0&0&0
\\ \noalign{\medskip}0&0&0&1&1&1&0&0&0\\ \noalign{\medskip}0&0&0&0&0&0
&1&1&1\end {array} \right] \mathbf{P}^{(2)}=
 \left[ \begin {array}{ccc|ccc|ccc} 1&0&0&1&0&0&1&0&0
\\ \noalign{\medskip}0&1&0&0&1&0&0&1&0\\ \noalign{\medskip}0&0&1&0&0&1
&0&0&1\end {array} \right] \mathbf{P}^{(2)}.
\end{equation}
Dashed vertical lines illustrate partitioning of matrices $\mathbf{R}^{(3)}$ 
and $\mathbf{L}^{(3)}$ into blocks of $\mathbf{I}$ 
and $\mathbf{J}$ type.

We can now make two remarks about matrices $\mathbf{R}^{(k)}$ 
and $\mathbf{L}^{(k)}$. First of all, using eq. (\ref{matrixcons}) recursively,
we can express every $\mathbf{P}^{(m)}$ for $m\in[1,k)$ by
$\mathbf{P}^{(k)}$,
\begin{equation}\label{downreductionmatrixform}
 \mathbf{P}^{(m)}=\left(\prod_{i=m+1}^k \mathbf{L}^{(i)} \right) \mathbf{P}^{(k)}.
\end{equation}
In the above, one could replace all (or only some) $\mathbf{L}$'s by $\mathbf{R}$'s, and the
equation would remain valid.

Secondly, note that both $\mathbf{L}^{(1)}$ and $\mathbf{R}^{(1)}$  
are single row matrices with all $N$ entries equal to 1. This implies that the product
$\mathbf{L}^{(1)} \mathbf{L}^{(2)}$ is a single row matrix with
all $N^2$ entries equal to 1, and, in general, for any $k\geq 1$,
\begin{equation}
 \prod_{i=1}^k \mathbf{L}^{(i)}=[\, \underbrace{1\,\, 1 \ldots 1}_{N^k}\, ].
\end{equation}
Again, one could replace here all (or some) $\mathbf{L}$'s by $\mathbf{R}$'s, and the
equation would remain valid. As a consequence of this, normalization 
condition (\ref{cons2}) can be written as $\mathbf{L}^{(1)} \mathbf{P}^{(1)}=1$,
or, replacing $\mathbf{P}^{(1)}$ by $\mathbf{L}^{(2)} \mathbf{P}^{(2)}$, as
$\mathbf{L}^{(1)} \mathbf{L}^{(2)} \mathbf{P}^{(2)}=1$, etc. In general, we can
write the normalization condition in the form
\begin{equation}
\left( \prod_{i=1}^k \mathbf{L}^{(i)} \right) \mathbf{P}^{(k)}=1,
\end{equation}
which, of course, is equivalent to
\begin{equation}\label{normforpk}
\sum_{i=1}^{N^k} \mathbf{P}^{(k)}_i=1. 
\end{equation}
Naturally, this was to be expected, since it is a consequence of measure additivity
and the fact that 
\begin{equation}
\bigcup_{\mathbf{b} \in {\mathcal{A}}^k} [\mathbf{b}]_i = X.
\end{equation}

After making the above remarks about consistency conditions and their matrix
form, let us turn our attention to the following problem.
In order to fully describe a shift-invariant probability measure
one needs to know all block probabilities $\mathbf{P}^{(i)}$ with $i=1,2,\ldots$,
and make sure that they satisfy consistency conditions. In practical
applications, however, it is often impossible to know \emph{all}
block probabilities, and instead one considers only
truncated sequence of block probabilities $\mathbf{P}^{(i)}$ for
$i=1,2,\ldots, k$. It is then important to know how many of these
are truly independent? The next proposition answers this question.
\begin{prop}\label{propnrind}
Among all block probabilities constituting components of
 $\mathbf{P}^{(1)} , \mathbf{P}^{(2)} , \dots, \mathbf{P}^{(k)} $
only $(N-1)N^{k-1}$ are linearly independent.
\end{prop}
\emph{Proof:}
Let us first note that vector $\mathbf{P}^{(i)}$ has $N^i$ components. 
Collectively, in $\mathbf{P}^{(1)} , \mathbf{P}^{(2)} , \dots, \mathbf{P}^{(k)} $ we
have, therefore, $\sum_{i=1}^k N^i=(N^{k+1}-N)/(N-1)$ block probabilities.
However, since all $\mathbf{P}^{(i)}$, $i\in [1,k)$, can be expressed in terms of 
$\mathbf{P}^{(k)}$ with the help of eq. (\ref{downreductionmatrixform}),
we can treat all of $\mathbf{P}^{(1)} , \mathbf{P}^{(2)} , \dots$, $\mathbf{P}^{(k-1)}$
as dependent. This leaves us with  $\mathbf{P}^{(k)}$ with $N^k$ components. However,
we also have 
\begin{equation}\label{symeqfork}
 \mathbf{L}^{(k)} \mathbf{P}^{(k)} = \mathbf{R}^{(k)} \mathbf{P}^{(k)}.
\end{equation}
Matrices  in the above have $N^{k-1}$ rows, thus we have $N^{k-1}$ equations for
$N^k$ variables. Are they all these equations independent?
Both $L$ and $R$ have the property that sum of each of their columns is 1.
Thus if we add all equations of (\ref{symeqfork}), we obtain identity
$\sum \mathbf{P}^{(k)} = \sum \mathbf{P}^{(k)}$, meaning that the number of
independent equations in eq. (\ref{symeqfork}) is $N^{k-1}-1$.
All of this takes care of consistency conditions (\ref{cons1}), but we also need to
consider normalization condition (\ref{cons2}) which, as remarked earlier, can
be written in equivalent form as equation involving components of $\mathbf{P}^{(k)}$,
that is, eq. (\ref{normforpk}). This additional equation increases our previously
obtained number of independent equations back to $N^{k-1}$. In the end, the number of
independent block probabilities, equal to number of variables minus number of
independent equations,  is $N^k - N^{k-1}=(N-1)N^{k-1}$. $\square$

Once we know  how many  independent block probabilities are there, we 
can express the remaining block probabilities  in terms of them. We need to choose
which block probabilities we declare to be independent. The following proposition
describes a natural choice. Before we state it, we need to introduce some 
additional notation. As explained in the proof of Proposition \ref{propnrind},
in the system of equations $\mathbf{R}^{(k)}\mathbf{P}^{(k)}=\mathbf{L}^{(k)}\mathbf{P}^{(k)}$
only $N^{k-1}-1$ equations are independent. We can, therefore, remove one of them, for example,
the last equation, and replace it by normalization condition $\sum \mathbf{P}^{(k)}=1$.
This will result in 
\begin{equation}
 \mathbf{M}^{(k)} \mathbf{P}^{(k)} = \left[ \begin {array}{c} 0\\ \vdots \\  0
\\ 1\end {array} \right],
\end{equation}
where the matrix $\mathbf{M}^{(k)}$ has been obtained from $\mathbf{R}^{(k)} -\mathbf{L}^{(k)} $
by setting every entry in the last row of $\mathbf{R}^{(k)} -\mathbf{L}^{(k)} $ to 1.
Let us now partition $\mathbf{M}^{(k)}$ into two submatrices, so that  the first $N^k-N^{k-1}$ 
columns of it are called $\mathbf{A}^{(k)}$, and the remaining $N^{k-1}$ columns are called $\mathbf{B}^{(k)}$,
so that
\begin{equation}
 \mathbf{M}^{(k)} = [\mathbf{A}^{(k)} \mathbf{B}^{(k)}].
\end{equation}
If we recall definitions of $\mathbf{L}^{(k)}$ and  $\mathbf{R}^{(k)}$ in eqs. (\ref{defL}) and (\ref{defR}),
we can easily verify that
\begin{equation}
\mathbf{B}^{(k)}=\left[
 \begin {array}{rrrr} 
 -1&0&\cdots&0\\ 
 0&-1& & \\
 \vdots & &\ddots& \\
 1&1&1&1
 \end{array} \right], 
\end{equation}
so that $\mathbf{B}^{(k)}$ can be constructed from zero $N^{k-1} \times N^{k-1}$ matrix  by
placing $-1$'s on the diagonal, and then filling the last row with  1's. The structure
of matrix $\mathbf{A}^{(k)}$ is a bit more complicated,
\begin{equation}
 \mathbf{A}^{(k)} = [\mathbf{J}_{1} \mathbf{J}_{2} \ldots \mathbf{J}_{N-1}]+
[\,\underbrace{\mathbf{B}^{(k)}\,\,\, 
\mathbf{B}^{(k)} \ldots \mathbf{B}^{(k)}}_{N-1}\,],
\end{equation}
where, as already defined, $\mathbf{J}_{m}$ is an  $N^{k-1}\times N^{k-1}$ matrix in which $m$-th
row consist of all 1's, and all other entries are equal to 0.
\begin{prop} \label{hilowdependenceprop}
 Let $\mathbf{P}^{(k)}$ be partitioned into two subvectors, $\mathbf{P}^{(k)}=
(\mathbf{P}^{(k)}_{Top}$, $\mathbf{P}^{(k)}_{Bot})$, where $\mathbf{P}^{(k)}_{Top}$
contains first $N^k-N^{k-1}$ entries of $\mathbf{P}^{(k)}$, and
$\mathbf{P}^{(k)}_{Bot}$ the remaining $N^{k-1}$ entries. Then
\begin{equation}\label{depbyindep}
 \mathbf{P}^{(k)}_{Bot}=
\left[ \begin {array}{c} 0\\ \vdots \\  0
\\ 1\end {array} \right]
 - \left(\mathbf{B}^{(k)}\right)^{-1} \mathbf{A}^{(k)}\mathbf{P}^{(k)}_{Top}.
\end{equation}
\end{prop}
\emph{Proof:}
we want to solve 
\begin{equation}
[\mathbf{A}^{(k)} \mathbf{B}^{(k)}]   \left[ \begin {array}{c} \mathbf{P}^{(k)}_{Top}\\[0.5em]
\mathbf{P}^{(k)}_{Bot}
\end {array} \right] =\left[ \begin {array}{c} 0\\ \vdots \\  0
\\ 1\end {array} \right]
\end{equation}
for $\mathbf{P}^{(k)}_{Bot}$.
Denoting the vector on the right hand side by $\mathbf{c}$ and performing block multiplication we 
obtain $ \mathbf{A}^{(k)} \mathbf{P}^{(k)}_{Top}+ \mathbf{B}^{(k)} \mathbf{P}^{(k)}_{Bot}=\mathbf{c}$.
The matrix $ \mathbf{B}^{(k)}$ is always invertible, and has the property $ (\mathbf{B}^{(k)})^{-1} \mathbf{c}=\mathbf{c}$.
This leads to $\mathbf{P}^{(k)}_{Bot}= \mathbf{c}
 - \left(\mathbf{B}^{(k)}\right)^{-1} \mathbf{A}^{(k)}\mathbf{P}^{(k)}_{Top}$, as desired. $\square$
\begin{cor} \label{reductioncorollary}
Among  block probabilities constituting components of
 $\mathbf{P}^{(1)}$ , $\mathbf{P}^{(2)} , \dots, \mathbf{P}^{(k)} $,
we can treat first $N^{k}-N^{k-1}$ entries of $\mathbf{P}^{(k)}$ as independent variables.
Remaining components of $\mathbf{P}^{(k)}$ can be obtained by using eq. (\ref{depbyindep}),
while  $\mathbf{P}^{(1)} , \mathbf{P}^{(2)} , \dots, \mathbf{P}^{(k-1)}$ can be obtained by
eq. (\ref{downreductionmatrixform}).
\end{cor}
Representation of all blocks $\mathbf{P}^{(1)} , \mathbf{P}^{(2)} , \dots, \mathbf{P}^{(k)} $ by
first $N^{k}-N^{k-1}$ entries of $\mathbf{P}^{(k)}$ will be called \emph{long block representation}.
As an example of this, let us consider the case of ${\mathcal{A}}=\{0,1,2\}$ ($N=3$) and 
$\mathbf{P}^{(1)} , \mathbf{P}^{(2)}, \mathbf{P}^{(3)}$. We have $3^3-3^2=18$ independent
block probabilities, all of length $3$. These are
\begin{multline*}
\{P(000),P(001),P(002),P(010),P(011),P(012),P(020),P(021),\\P(022),
P(100),P(101),P(102),P(110),P(111),P(112),P(120),\\P(121),P(122)\}.
\end{multline*}
Remaining 21 block probabilities, expressible in terms of the above, are
\begin{multline*}
 \{P(200),P(201),P(202),P(210),P(211),P(212),P(220),P(221),\\P(222),
P(00),P(01),P(02),P(10),P(11),P(12),P(20),P(21),\\P(22),
P(0),P(1),P(2)\}.
\end{multline*}
Since there are there are total  $\sum_{i=1}^k N^i=(N^{k+1}-N)/(N-1)$ block probabilities in 
$\mathbf{P}^{(1)} , \mathbf{P}^{(2)} , \dots, \mathbf{P}^{(k)} $, the fraction
of independent block probabilities among all block probabilities up to length $k$
is
\begin{equation}
 \mathrm{Ind}(N,k):=\frac{(N-1) (N^k-N^{k-1})}{N^{k+1}-N}.
\end{equation}
For fixed $N$, $\mathrm{Ind}(N,k)$ decreases as a function of $k$, and tends to the limit
\begin{equation}
 \lim_{k \to \infty}\mathrm{Ind}(N,k)=\frac{(N-1)^2}{N^2}
\end{equation}
The above reaches minimum $1/4$ at $N=2$, thus $\mathrm{Ind}(N,k)>1/4$ for all $k\geq 1$, $N > 1$.
This means that  the long block expression is most ``economical'' for the binary alphabet. For
example, for $N=2$ and $k=3$, among $\mathbf{P}^{(1)} , \mathbf{P}^{(2)}, \mathbf{P}^{(3)}$ we have
only 4 independent blocks, $P(000),P(001),P(010)$ and $P(011)$. Remaining 10 probabilities
can be expressed as follows,
\begin{align} \label{longblockexamplek3}
 \left[ \begin {array}{c} P(100)\\
 P(101)\\ P(110)\\ 
P(111)\end {array} \right] &= \left[ \begin {array}{c} P(001)\\ 
-P(001)+P(010)+P(011)\\
 P(011)\\
1 -P(000)- P(001)-2\,P(010)-3\,P(011)\end {array} \right],  \nonumber \\
\left[ \begin {array}{c} P(00)\\ P(01)\\ P(10)\\P(11)
\end {array} \right] &= \left[ \begin {array}{c} 
P(000)+P(001)\\ 
P(010)+P(011)\\ 
P(010)+P(011)\\ 
1-P(000)-P(001)-2\,P(010)-2\,P(011)
\end {array} \right],  \nonumber \\
 \left[ \begin {array}{c} P(0)\\ P(1)
\end {array} \right] &= \left[ \begin {array}{c}
 P(000)+P(001)+P(010)+P(011)\\
 1-P(000)-P(001)-P(010)-P(011)\end {array}
 \right]. 
\end{align}

Of course, the long block representation is not the only one possible. 
We will describe below yet another representation, which
is in some sense complementary to the the long block one. It declares
as independent blocks of shortest possible length, thus it will be called
\emph{short block representation}.

\begin{figure}
{
 \small
\setlength{\fboxsep}{0.2em}
\begin{equation*}
\xymatrix @R=0.0em{
k=1 & k=2 & k=3 &  \\ 
 & & & \\
\fbox{P(0)} \ar@{|->}[r]^{-}  &     \fbox{P(00)}\ar[ddd]^{-} \ar @{|->}[r]^{-}   &       \fbox{ P(000)} \ar @/^2.5pc/[d]^{-} & \mbox{\hspace{4em}}\ar@{>-<}[ddddddddddddddddd]^{
\substack{\displaystyle \,\, N^k-N^{k-1} \mathrm{\,\,entries}\\[1em] \displaystyle \,\,\mathbf{P}^{(k)}_{Top} }}\\
                       &                                           &       \fbox{ P(001)} \ar @/^2.5pc/[d]^{=} & \\
                       &                                           &       \underline{P(002)}  & \\
                       &     \fbox{P(01)}\ar[ddd]^{=}  \ar @{|->}[r]^{-}  &        \fbox{ P(010)} \ar @/^2.5pc/[d]^{-} & \\
                       &                                           &        \fbox{ P(011)} \ar @/^2.5pc/[d]^{=} & \\
                       &                                           &       \underline{P(012)}  & \\
                       &     \underline{P(02)}             \ar @{|->}[r]^{-}   &        \fbox{ P(020)} \ar @/^2.5pc/[d]^{-} & \\
                       &                                           &        \fbox{ P(021)} \ar @/^2.5pc/[d]^{=} & \\
                       &                                           &       \underline{P(022)}  & \\
\fbox{P(1)} \ar@{|->}[r]^{-}  &     \fbox{P(10)}\ar[ddd]^{-}  \ar @{|->}[r]^{-}  &        \fbox{ P(100)} \ar @/^2.5pc/[d]^{-} & \\
                       &                                           &        \fbox{ P(101)} \ar @/^2.5pc/[d]^{=} & \\
                       &                                           &       \underline{P(102)}  & \\
                       &     \fbox{P(11)}\ar[ddd]^{=} \ar @{|->}[r]^{-}   &        \fbox{ P(110)} \ar @/^2.5pc/[d]^{-} & \\
                       &                                           &        \fbox{ P(111)} \ar @/^2.5pc/[d]^{=} & \\
                       &                                           &       \underline{P(112)}  & \\
                       &     \underline{P(12)}\ar @{|->}[r]^{-}                &        \fbox{ P(120)} \ar @/^2.5pc/[d]^{-} & \\
                       &                                           &        \fbox{ P(121)} \ar @/^2.5pc/[d]^{=} & \\
                       &              &       \underline{P(122)} & \\
P(2)                   &     P(20)    &       P(200) & \ar@{>-<}[dddddddd]^{\substack{\displaystyle  N^k \mathrm{\,\,entries} \\[1em] 
\displaystyle  \,\,\mathbf{P}^{(k)}_{Bot}= \mathbf{c}
 - \left(\mathbf{B}^{(k)}\right)^{-1} \mathbf{A}^{(k)}\mathbf{P}^{(k)}_{Top}  }} \\
                       &              &       P(201) & \\
                       &              &       P(202) & \\
                       &     P(21)    &       P(210) & \\
                       &              &       P(211) & \\
                       &              &       P(212) & \\
                       &     P(22)    &       P(220) & \\
                       &              &       P(221) & \\
                       &              &     	 P(222) &
}
\end{equation*}
} 
\caption{Generation of short block representation for $N=3$ and $\mathbf{P}^{(k)}$ for $k=1,2,3$. Independent block probabilities are boxed, while 
dependent block probabilities obtained from probabilities of shorter blocks are underlined.}\label{FigGenCan}
\end{figure}
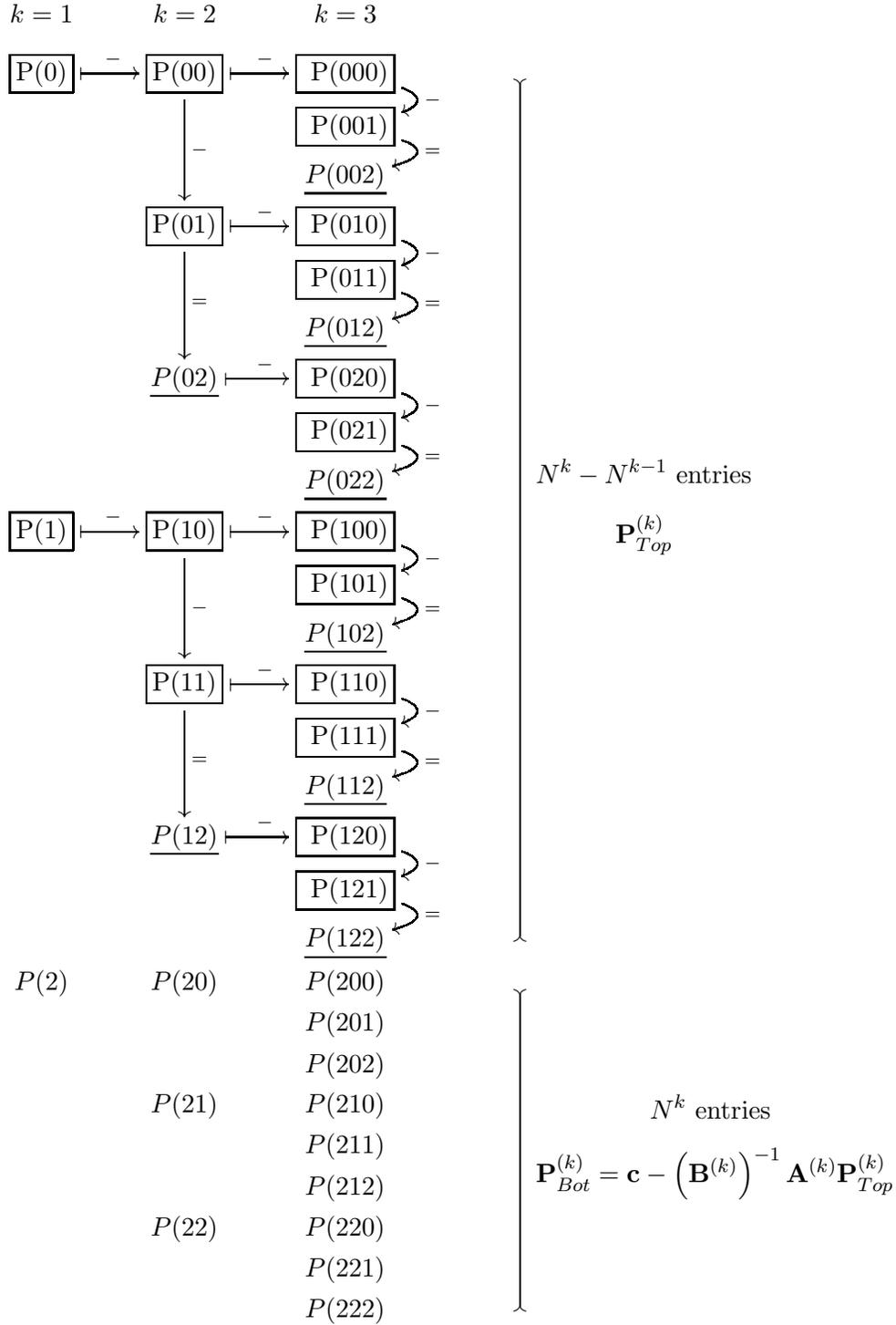

It is constructed as follows. We start, as before,  with block probabilities
$\mathbf{P}^{(1)} , \mathbf{P}^{(2)} , \dots, \mathbf{P}^{(k)} $, and we 
arrange each of the vectors $\mathbf{P}^{(i)}$ in a vertical column.
Example of this is shown in Figure \ref{FigGenCan}. In each vector $\mathbf{P}^{(i)}$,
we put aside last $N^{i-1}$ entries, and in what remains, we underline every $N$-th entry,
starting from the top. Entries which are still left are framed (cf.  Figure \ref{FigGenCan}), and those we declare to
be independent. It is straightforward to verify that we have $N^{k}-N^{k-1}$ independent entries, as we should.
Now how do we express dependent entries in terms of independent ones? 
In each vector, starting from the left, we replace each underlined entry by a linear combinations of boxed entries
from the same column and (possibly) entries from the column on the left hand side, 
by following the path which starts with  $\longmapsto$ arrow and which ends at the underlined entry in question.
For example, for $P(02)$, such path is $P(0)\longmapsto P(00) \to P(01) \to P(02)$. Labels above
arrows indicate how the equation is to be constructed, in this case
\begin{equation}
P(0) - P(00) - P(01) = P(02).
 \end{equation}
All arrows are labeled with ``$-$'', except those which point toward underlined entries, which are labeled with ``$=$''.

Once we are done with all underlined entries in a given vector, we express all entries marked as $\mathbf{P}^{(k)}_{Bot}$
 by $\mathbf{P}^{(k)}_{Top}$,
using eq. (\ref{depbyindep}). We then move to the next vector on the right and repeat the same procedure, until
all vectors are dealt with. By inspecting Figure \ref{FigGenCan}, the reader can verify that
the short block representation utilizes short blocks as much possible, and that, in fact, it is not possible
to declare a larger number of short blocks as independent.

In order to describe the above algorithm in a more formal way, let us define vector of
\emph{admissible entries} for short block representation, $\mathbf{P}^{(k)}_{adm}$, as follows.
Let us take vector $\mathbf{P}^{(k)}$ in which block probabilities are arranged in lexicographical order,
indexed by an index $i$ which runs from 1 to $N^k$. Vector $\mathbf{P}^{(k)}_{adm}$  consists of
all entries of $\mathbf{P}^{(k)}$ for which the index $i$ is not divisible by $N$ and for which
$i<N^k-N^{k-1}$. For example, for $N=3$ and $k=2$ we have
\begin{equation*}\mathbf{P}^{(2)}=
[P(00),P(01),P(02),P(10),P(11),P(12),P(20),P(21),P(22)]^T,
\end{equation*}
and we need to select entries with $i$ not divisible by 3 and $i<6$, which leaves $i=1,2,4,5$, hence
$$\mathbf{P}^{(2)}_{adm}=[P(00),P(01),P(10),P(11)]^T.$$

Vector of independent block probabilities in \emph{short block representation} is now defined as
\begin{equation}
 \mathbf{P}^{(k)}_{short}
 =\left[ \begin{array}{c} \mathbf{P}^{(1)}_{adm}\\[0.2em] \mathbf{P}^{(2)}_{adm} \\ \vdots \\  \mathbf{P}^{(k)}_{adm}
\end{array} \right].
\end{equation}
For $N=3$ and $k=2$, elements of $\mathbf{P}^{(3)}_{short}$ are shown in Figure 1 in red color. Note that
the length of $\mathbf{P}^{(k)}_{short}$ is the same as $\mathbf{P}^{(k)}_{Top}$. We can, therefore,
transform one into the other by a linear transformation. The form of this transformation can be 
deduced from Figure~1. Consider, for example, $k=2$, so that $\mathbf{P}^{(2)}_{Top}=[P(00),P(01),P(02),P(10),P(11),P(12)]^T$
and 
$\mathbf{P}^{(2)}_{adm}=[P(00),P(01),P(10),P(11)]^T$, $\mathbf{P}^{(1)}_{Bot}=\mathbf{P}^{(1)}_{adm}=[P(0),P(1)]^T$.
From Figure 1, we read
\begin{align}
 P(00)&=P(00), \nonumber \\
 P(01)&=P(01), \nonumber \\
 P(02)&=P(0)-P(00)-P(01), \nonumber \\
 P(10)&=P(00), \nonumber \\
 P(11)&=P(01), \nonumber \\
 P(12)&=P(1)-P(10)-P(11), 
\end{align}
where, if an element $P(\mathbf{b})$ of $\mathbf{P}^{(2)}_{Top}$ was admissible, we wrote $P(\mathbf{b})=P(\mathbf{b})$,
and if it was admissible, we expressed it in terms of probabilities of shorter blocks. The above can be written as
\begin{equation}
 \mathbf{P}^{(2)}_{Top}= \left[
                        \begin{array}{cc}
                         0 & 0\\
                         0 & 0\\
                         1 & 0\\
                         0 & 0\\
                         0 & 0\\
                         0 & 1
                        \end{array} \right]  \mathbf{P}^{(1)}_{adm} +
                    \left[ \begin{array}{rrrr}
                         1 & 0 & 0 & 0\\
                         0 & 1 & 0 & 0\\
                         -1 & -1 & 0 & 0\\
                         0 & 0 & 1 & 0\\
                         0 & 0 & 0 & 1\\
                         0 & 0 & -1 & -1\\
                        \end{array} \right]  \mathbf{P}^{(2)}_{adm}.
\end{equation}
This expresses $\mathbf{P}^{(2)}_{Top}$ in terms of $\mathbf{P}^{(1)}_{adm}$ and $\mathbf{P}^{(2)}_{adm}$,
that is, in terms of $\mathbf{P}^{(2)}_{short}$. One can similarly show that for general
$k>1$,
\begin{equation} \label{recursiveshort}
\mathbf{P}^{(k)}_{Top}=\mathbf{C}^{(k)}  \mathbf{P}^{(k-1)}_{Top} +\mathbf{D}^{(k)} \mathbf{P}^{(k)}_{adm},
\end{equation}
where
\begin{equation}
 \mathbf{C}^{(k)}= \mathrm{diag} (\underbrace{\mathbf{e}_{N}, \mathbf{e}_{N}, \ldots, \mathbf{e}_{N}}_{N^{k-1}-N^{k-2}} ),\,\,\,\,
\mathbf{e}_{N}=\left. \left[ \begin {array}{c} 0\\ \vdots \\  0
\\ 1\end {array} \right] \right\}N
\end{equation}
\begin{equation}
 \mathbf{D}^{(k)}= \mathrm{diag} (\underbrace{\mathbf{D}_{N}, \mathbf{D}_{N}, \ldots, \mathbf{D}_{N}}_{N^{k-1}-N^{k-2}} ),\,\,\,\,
\mathbf{D}_{N}= \underbrace{\left[ \begin {array}{c} \mathbf{I}_{N-1} \\
 -1,-1,\ldots,-1\end {array} \right]}_{N-1}. 
\end{equation}
Note that 
$\mathbf{C}^{(k)}$ has $N^k-N^{k-1}$ rows and $N^{k-1}-N^{k-2}$ columns, while 
$\mathbf{D}^{(k)}$ has $N^k-N^{k-1}$ rows and $(N^{k-1}-N^{k-2})(N-1)$ columns.	
Applying eq. (\ref{recursiveshort}) $k-1$ times recursively, one obtains, for $k>2$,
\begin{equation}  \label{solofrecursionk}
\mathbf{P}^{(k)}_{Top}=\mathbf{C}^{(k)}\mathbf{C}^{(k-1)}\ldots\mathbf{C}^{(2)}\mathbf{P}^{(1)}_{adm}
+\sum_{i=2}^{k-1} \mathbf{C}^{(k)}\mathbf{C}^{(k-1)}\ldots\mathbf{C}^{(i+1)} \mathbf{D}^{(i)} \mathbf{P}^{(i)}_{adm}
+\mathbf{D}^{(k)}\mathbf{P}^{(k)}_{adm}.
\end{equation}
When $k=2$ no recursion is needed, as eq. (\ref{recursiveshort})  becomes
\begin{equation} \label{solofrecursion2}
\mathbf{P}^{(2)}_{Top}=\mathbf{C}^{(2)}  \mathbf{P}^{(1)}_{adm} +\mathbf{D}^{(2)} \mathbf{P}^{(2)}_{adm},
\end{equation}
and for $k=1$ we simply have 
\begin{equation} \label{solofrecursion1}
\mathbf{P}^{(1)}_{Top}=\mathbf{P}^{(1)}_{adm}=\mathbf{P}^{(1)}_{short}. 
\end{equation}
If we define
\begin{equation}
 \mathbf{M}^{(k)}_{short}=
\begin{cases}
\Big[\mathbf{C}^{(k)}\mathbf{C}^{(k-1)}\ldots\mathbf{C}^{(2)},
 \underbrace{\mathbf{C}^{(k)}\mathbf{C}^{(k-1)}\ldots\mathbf{C}^{(i+1)} \mathbf{D}^{(i)}}_{\mathrm{repeated\,\, for}\,\,i=2\ldots k-1 },
\mathbf{D}^{(k)} \Big], & k>2 \\
 \left[\mathbf{C}^{(2)}, \mathbf{D}^{(2)}\right], & k=2\\
\mathbf{I}_{N-1}, & k=1
\end{cases}
\end{equation}
then equations (\ref{solofrecursionk}--\ref{solofrecursion1}) can be written as
\begin{equation} \label{shorttolongtrafo}
 \mathbf{P}^{(k)}_{Top}=\mathbf{M}^{(k)}_{short} \mathbf{P}^{(k)}_{short}.
\end{equation}
\begin{prop} \label{shortblockprop}
Among  block probabilities constituting components of
 $\mathbf{P}^{(1)}$, $\mathbf{P}^{(2)} , \dots, \mathbf{P}^{(k)} $,
we can treat entries of $\mathbf{P}^{(k)}_{short}$ as independent variables.
One can express first $N^k-N^{k-1}$ components of $\mathbf{P}^{(k)}$ by $\mathbf{P}^{(k)}_{short}$
by means of eq. (\ref{shorttolongtrafo}). Remaining components of $\mathbf{P}^{(k)}$ can be obtained by using eq. (\ref{depbyindep}),
while  $\mathbf{P}^{(1)} , \mathbf{P}^{(2)} , \dots, \mathbf{P}^{(k-1)}$ can be obtained by
eq. (\ref{downreductionmatrixform}).
\end{prop}
Let us now apply the procedure described above to the $N=2$ and $k=3$ case, the same as
we already considered for long block representation. Among components of $\mathbf{P}^{(1)} , \mathbf{P}^{(2)}$ and  
$\mathbf{P}^{(3)}$ we have only four independent block probabilities,
$\mathbf{P}^{(3)}_{short}=[P(0), P(00), P(000), P(010)]^T$, and 10 dependent probabilities.
We first partition $\mathbf{P}^{(3)}$ into two subvectors, $\mathbf{P}^{(3)}_{Top}=[P(000)$, $P(001),P(010),P(011)]^T$
and $\mathbf{P}^{(3)}_{Bot}=[P(100),P(101),P(110),P(111)]^T$.
 Eq. (\ref{shorttolongtrafo}) takes the form
\begin{equation} \label{eqp3ind}
\mathbf{P}^{(3)}_{Top}=\left[ \begin {array}{l}
        P(000)\\
        P(001)\\
        P(010)\\
        P(011)
       \end {array}
\right] = \mathbf{M}^{(3)}_{short} \mathbf{P}^{(3)}_{short}=
 \left[ \begin {array}{cccc} 0&0&1&0\\ \noalign{\medskip}0&1&-1&0
\\ \noalign{\medskip}0&0&0&1\\ \noalign{\medskip}1&-1&0&-1\end {array}
 \right] 
\left[ \begin {array}{l}
        P(0)\\
        P(00)\\
        P(000)\\
        P(010)
       \end {array}
\right].
\end{equation}
Components of $\mathbf{P}^{(3)}_{Bot}$ can be obtained from eq. (\ref{depbyindep}),
\begin{equation} \label{eqp3dep}
 \left[ \begin {array}{l}
        P(100)\\
        P(101)\\
        P(110)\\
        P(111)
       \end{array}\right]=
 \left[ \begin {array}{l}
         0\\
         0\\
         0\\
         1
        \end {array} \right] -
  \left[ \begin {array}{cccc} 0&-1&0&0\\ \noalign{\medskip}0&1&-1&-1
\\ \noalign{\medskip}0&0&0&-1\\ \noalign{\medskip}1&1&2&3\end {array}
 \right] 
\left[ \begin {array}{l}
        P(000)\\
        P(001)\\
        P(010)\\
        P(011)
       \end {array}
\right],
\end{equation}
and we can use eq. (\ref{eqp3ind}) again to replace $[P(000),P(001),P(010),P(011)]^T$ on the right hand side
by $\mathbf{M}^{(3)}_{short} \mathbf{P}^{(3)}_{short}$.
Equations (\ref{eqp3ind}) and (\ref{eqp3dep}), therefore, yield all components of $\mathbf{P}^{(3)}$. By applying eq.
(\ref{downreductionmatrixform}) we can obtain $\mathbf{P}^{(2)}$ and of $\mathbf{P}^{(1)}$.
This will yield the following 10 dependent blocks probabilities
expressed in terms of  elements of $\mathbf{P}^{(3)}_{short}$,
\begin{align} \label{shortform3}
\left[ \begin {array}{c} 
P(001)\\
P(011)\\  
P(100) \\
P(101)\\
P(110)\\
P(111)
\end {array} \right] &= 
\left[ \begin {array}{c} 
P(00)-P(000) \\ 
P(0)-P(00) -P(010) \\  
P(00)-P(000) \\  
P(0)-2 P(00)+P(000) \\ 
P(0)-P(00)-P(010)\\  
 1-3 P(0) +2 P(00) +P(010) 
\end {array} \right]. \nonumber \\
\left[ \begin {array}{c} 
P(01)\\
P(10)\\  
P(11)
\end {array} \right] &= 
\left[ \begin {array}{c} 
P(0) -P(00) \\
P(0) -P(00) \\ 
 1-2 P(0)+P(00)
\end {array} \right], \nonumber  	\\
 P(1) &= 1-P(0). 
\end{align}
We can see that the resulting expressions are shorter than in the case of long block representation given
by eq. (\ref{longblockexamplek3}). 
Indeed, the short block representation is more ``natural'' and often helps to gain insight into the 
properties of the probability measure it describes. We will see this when this representation is used to simplify 
local structure theory equations.
\section{Bayesian Extension}
From what we have seen so far, it is clear that the knowledge of $\mathbf{P}^{(k)}$ is
enough to determine all $\mathbf{P}^{(i)}$ with $i<k$. What about $i>k$? Obviously, since
the number of independent components in $\mathbf{P}^{(i)}$ is greater than in $\mathbf{P}^{(k)}$
for $i>k$, there is no hope to determine $\mathbf{P}^{(i)}$ using only $\mathbf{P}^{(k)}$.
It is possible, however, to approximate longer block probabilities by shorter block
probabilities using the idea of Bayesian extension.

Suppose now that we want to approximate $P(a_1a_2\ldots a_{k+1})$ by 
$P(a_1a_2$ $\ldots a_{k})$. One can say that by knowing $P(a_1a_2\ldots a_{k})$ we know how
values of individual symbols in a block are correlated providing that symbols are not farther apart
than $k-1$. We do not know, however, anything about correlations on the larger length scale.
The only thing we can do in this situation is to simply neglect these higher length correlations,
and assume that if a block of length $k$ is extended by adding another symbol to it on the right, then
the the conditional probability of finding a particular value of that symbol does not significantly
depend on the left-most symbol, i.e.,
\begin{equation}
\frac{P(a_1a_2\ldots a_{k+1})}{P(a_1\ldots a_{k})} \approx \frac{ P(a_{2}\ldots a_{k+1})  }
{P(a_2 \ldots a_{k})}.
\end{equation}
This produces the desired approximation of $k+1$ block probabilities by $k$-block and $k-1$ block probabilities,
\begin{equation} \label{bayapprox}
 P(a_1a_2\ldots a_{k+1}) \approx \frac{ P(a_1\ldots a_{k}) P(a_{2}\ldots a_{k+1}) }
{P(a_2 \ldots a_{k})},
\end{equation}
where we assume that the denominator is positive. If the denominator is zero, then
we take  $P(a_1a_2\ldots a_{k+1}) =0$.
In order to avoid writing separate cases for denominator equal to zero, we define  ``thick bar'' fraction as
\begin{equation} \label{convention1}
 \mfrac{a}{b}:=\begin{cases}
               {\displaystyle \frac{a}{b}} & \mathrm{if\,\,} b \neq 0\\[1em]
                   0 & \mathrm{if\,\,} b = 0.      
                     \end{cases}
\end{equation}
Note that eq. (\ref{bayapprox}) only makes sense if $k>1$. For $k=1$, the approximation
is
\begin{equation}
 P(a_1a_2) \approx P(a_1) P(a_2).
\end{equation}
Again, in order to avoid writing the $k=1$ case separately, we adopt notational convention that 
\begin{equation} \label{convention2}
 P(a_m \ldots a_n) =1 \mathrm{\,\, whenever \,\,} n>m,
\end{equation}
and then eq. (\ref{bayapprox}) remains valid even for $k=1$.
Using notational conventions given in eq. (\ref{convention1} ) and (\ref{convention2})
and applying our approximation recursively $m$ times we can express $k+m$ block
probabilities in terms of $k$ and $k-1$-block probabilities,
\begin{equation}
 {P}(a_1a_2 \ldots a_{k+m}) \approx \mfrac
{ 
\prod_{i=1}^{m+1} P(a_i \ldots a_{i+k-1})  
}
{ 
\prod_{i=1}^{m} P(a_{i+1} \ldots a_{i+k-1}) 
}.
\end{equation}
Note that if we want, we can write the right hand side of the above in terms of only $k$-block
probabilities, by substituting  in the denominator 
\begin{equation}
P(a_{i+1} \ldots a_{i+k-1})=\sum_{b \in {\mathcal{A}}} P(a_{i+1} \ldots a_{i+k-1}b).
\end{equation}
%
%
%
%
%
%
%
%
\begin{prop}
Let $\mu \in \mathfrak{M}_{\sigma}(X)$ be a measure with associated block probabilities $P: {\mathcal{A}}^{\star} \to [0,1]$,
$P(\mathbf{b})=\mu([\mathbf{b}]_i)$ for all $i\in \mathbb{Z}$ and $\mathbf{b}\in {\mathcal{A}}^{\star}$.
For $k>0$,  define
$\widetilde{P}: {\mathcal{A}}^{\star} \to [0,1]$ such that 
\begin{equation}
 \widetilde{P}(a_1a_2 \ldots a_{p})=
\begin{cases}
\,\,\,\,\,\,\,P(a_1a_2 \ldots a_{p}) & \mathrm{if\,\,} p \leq k ,\\[0.5em]
\displaystyle  \mfrac
{ 
\prod_{i=1}^{p-k+1} P(a_i \ldots a_{i+k-1})  
}
{ 
\prod_{i=1}^{p-k}  P(a_{i+1} \ldots a_{i+k-1}) 
}  & \mathrm{otherwise}.        
\end{cases}
\end{equation}
Then $\widetilde{P}$ determines a shift-invariant probability measure $\widetilde{\mu}^{(k)} \in \mathfrak{M}_\sigma(X)$,
 to
be called \emph{Bayesian approximation of $\mu$ of order $k$}. 
\end{prop}
\emph{Proof.}
If  $\mathbf{b}=b_1b_2\ldots b_n$, we will denote subblocks of $\mathbf{b}$ by 
$\mathbf{b}_{[i,j]}=b_ib_{i+1}\ldots b_j$.
Using Theorem \ref{extensionfromblocks}, all we need to do is to show that conditions (\ref{cons1}) and (\ref{cons2})
are satisfied by $\widetilde{P}$. 
The second one holds for $\widetilde{P}$, because it obviously holds
for $P$. For the same reason eq. (\ref{cons1}) holds for block $\mathbf{b}$ of length up to $k-1$.
For $\mathbf{b}=b_1b_2\ldots b_p$, $p \geq k$, we have
\begin{multline}
\sum_{a \in {\mathcal{A}}} \widetilde{P}(\mathbf{b}a)=\sum_{a \in {\mathcal{A}}} 
  \mfrac
{ 
 \prod_{i=1}^{p-k+1} P(\mathbf{b}_{[i,i+k-1]}) \cdot P(\mathbf{b}_{[p-k+2,p]} a) 
}
{ 
 \prod_{i=1}^{p-k}  P(\mathbf{b}_{[i+1,i+k-1]}) \cdot P(\mathbf{b}_{[p-k+2,p]} ) 
}
\\=
  \mfrac
{ 
\prod_{i=1}^{p-k+1} P(\mathbf{b}_{[i,i+k-1]}) 
}
{ 
 \prod_{i=1}^{p-k}  P(\mathbf{b}_{[i+1,i+k-1]}) \cdot P(\mathbf{b}_{[p-k+2,p]} ) 
} 
\sum_{a \in {\mathcal{A}}} P(\mathbf{b}_{[p-k+2,p]} a)
\\= \mfrac
{ 
\prod_{i=1}^{p-k+1} P(\mathbf{b}_{[i,i+k-1]}) 
}
{ 
 \prod_{i=1}^{p-k}  P(\mathbf{b}_{[i+1,i+k-1]}) \cdot P(\mathbf{b}_{[p-k+2,p]} ) 
} P(\mathbf{b}_{[p-k+2,p]} )
=\widetilde{P}(\mathbf{b}).
\end{multline}
One can similarly prove that $\sum_{a \in {\mathcal{A}}} \widetilde{P}(a\mathbf{b})=\widetilde{P}(\mathbf{b})$. $\square$

When there exists $k$ such that Bayesian approximation of $\mu$ of order $k$ is equal to $\mu$, we call $\mu$
a \emph{Markov measure} or a \emph{finite block measure} of order $k$. The space of Markov measures of order $k$
will be denoted by $\mathfrak{M}^{(k)}(X)$,
\begin{equation}
 \mathfrak{M}^{(k)}(X)=\{\mu \in \mathfrak{M}_{\sigma}(X): \mu=\widetilde{\mu}^{(k)}\}.
\end{equation}

It is often said that the Bayesian approximation ``maximizes entropy''. In order to state this property 
in a formal way, let us define  \emph{entropy density} of shift-invariant measure  $\mu \in \mathfrak{M}_\sigma(X) $ as
\begin{equation}
h(\mu)= \lim_{n \to \infty} - \frac{1}{n} \sum_{\mathbf{b} \in {\mathcal{A}}^n} P(\mathbf{b}) \log P(\mathbf{b}),
\end{equation}
where, as usual, $P(\mathbf{b})=\mu([\mathbf{b}]_i)$ for all $i\in \mathbb{Z}$ and $\mathbf{b}\in {\mathcal{A}}^{\star}$.
The following two propositions and the main ideas behind their  proofs are due to M. Fannes and A. Verbeure \cite{Fannes84}.
\begin{prop}
For any  $\mu \in \mathfrak{M}_\sigma(X) $, the entropy density of the $k$-th order Bayesian approximation of $\mu$
is given by
\begin{equation}
h( \widetilde{\mu}^{(k)})=
\sum_{\mathbf{a} \in {\mathcal{A}}^{k-1}} P(\mathbf{a}) \log P(\mathbf{a})
-\sum_{\mathbf{a} \in {\mathcal{A}}^k} P(\mathbf{a}) \log P(\mathbf{a}).
\end{equation}
\end{prop}
\emph{Proof:}
Since we are interested in $n \to \infty$ limit, let us consider $n>k$. Then
\begin{multline} \label{entropy1}
 \sum_{\mathbf{b} \in {\mathcal{A}}^n} \widetilde{P}(\mathbf{b}) \log \widetilde{P}(\mathbf{b})
=\sum_{\mathbf{b} \in {\mathcal{A}}^n} \widetilde{P}(\mathbf{b})
\log
\mfrac
{ 
\prod_{i=1}^{n-k+1} P(\mathbf{b}_{[i,i+k-1]})  
}
{ 
\prod_{i=1}^{n-k}  P(\mathbf{b}_{[i+1,i+k-1]}) 
} 
\\=
\sum_{\mathbf{b} \in {\mathcal{A}}^n} \widetilde{P}(\mathbf{b})
\sum_{i=1}^{n-k+1} \log P(\mathbf{b}_{[i,i+k-1]})  
-
\sum_{\mathbf{b} \in {\mathcal{A}}^n} \widetilde{P}(\mathbf{b})
\sum_{i=1}^{n-k}  \log P(\mathbf{b}_{[i+1,i+k-1]}) 
\end{multline}
For any $i\in [1,n-k+1]$,
\begin{multline}
 \sum_{\mathbf{b} \in {\mathcal{A}}^n} \widetilde{P}(\mathbf{b})
 \log P(\mathbf{b}_{[i,i+k-1]})  = \\
\sum_{\substack{\mathbf{b}_{[1,i-1]} \\ \in {\mathcal{A}}^{i-1} }}
\sum_{\substack{\mathbf{b}_{[i,i+k-1]}\\ \in {\mathcal{A}}^{k} }}
\sum_{\substack{\mathbf{b}_{[i+k,n]}\\ \in {\mathcal{A}}^{n-i-k+1} }} 
\widetilde{P}( \mathbf{b}_{[1,i-1]} \mathbf{b}_{[i,i+k-1]} \mathbf{b}_{[i+k,n]})
 \log P(\mathbf{b}_{[i,i+k-1]})\\ = \sum_{\mathbf{a} \in {\mathcal{A}}^k} \widetilde{P}(\mathbf{a}) 
\log P(\mathbf{a})
=\sum_{\mathbf{a} \in {\mathcal{A}}^k} P(\mathbf{a}) \log P(\mathbf{a}),
\end{multline}
and, by the same reasoning, any $i\in [1,n-k]$,
\begin{equation}
 \sum_{\mathbf{b} \in {\mathcal{A}}^n} \widetilde{P}(\mathbf{b}) \log P(\mathbf{b}_{[i+1,i+k-1]})=
\sum_{\mathbf{a} \in {\mathcal{A}}^{k-1}} P(\mathbf{a}) \log P(\mathbf{a}).
\end{equation}
Using this, eq. (\ref{entropy1}) becomes
\begin{multline} \label{pblogpb}
\sum_{\mathbf{b} \in {\mathcal{A}}^n} \widetilde{P}(\mathbf{b}) \log \widetilde{P}(\mathbf{b})=\\
(n-k+1)\sum_{\mathbf{a} \in {\mathcal{A}}^k} P(\mathbf{a}) \log P(\mathbf{a}) 
-
(n-k) \sum_{\mathbf{a} \in {\mathcal{A}}^{k-1}} P(\mathbf{a}) \log P(\mathbf{a}).
\end{multline}
Dividing this by $-n$ and taking the limit $n \to \infty$ we obtain the desired expression. $\square$

\begin{thm}
 For any  $\mu \in \mathfrak{M}_\sigma(X) $ and any $k>0$, the entropy density of $\mu$ does not exceed the
entropy density of its $k$-th order Bayesian approximation,
\begin{equation} \label{entropyineq}
 h(\mu) \leq h(\widetilde{\mu}^{(k)}).
\end{equation}
\end{thm}
\emph{Proof:}
Let 
\begin{align}
H_n(\mu)&=-\sum_{\mathbf{b} \in {\mathcal{A}}^n}  P(\mathbf{b}) \log P(\mathbf{b}),\\
H_n(\widetilde{\mu}^{(k)})&=-\sum_{\mathbf{b} \in {\mathcal{A}}^n}  \widetilde{P}(\mathbf{b}) \log \widetilde{P}(\mathbf{b}),\\
\end{align}
We will use convexity of $f(x)=x \log x$,
\begin{equation}
x \log x - y \log y \leq (x-y)(1+\log x).
\end{equation}
Applying this inequality to $S_n(\mu)-S_n(\widetilde{\mu}^{(k)})$ for $n>k$ we obtain
\begin{multline}
 H_n(\mu)-H_n(\widetilde{\mu}^{(k)})=\sum_{\mathbf{b} \in {\mathcal{A}}^n}  \widetilde{P}(\mathbf{b}) \log \widetilde{P}(\mathbf{b})
-\sum_{\mathbf{b} \in {\mathcal{A}}^n}  P(\mathbf{b}) \log P(\mathbf{b})\\
\leq 
\sum_{\mathbf{b} \in {\mathcal{A}}^n} \left(\widetilde{P}(\mathbf{b}) - {P}(\mathbf{b}) \right)
\left(1+ \log \widetilde{P}(\mathbf{b})\right)\\
=
\sum_{\mathbf{b} \in {\mathcal{A}}^n} \widetilde{P}(\mathbf{b}) \log \widetilde{P}(\mathbf{b}) - 
\sum_{\mathbf{b} \in {\mathcal{A}}^n}
{P}(\mathbf{b})  \log \widetilde{P}(\mathbf{b}) ,
\end{multline}
where we used the fact that $\sum_{\mathbf{b} \in {\mathcal{A}}^n} \widetilde{P}(\mathbf{b}) = 
\sum_{\mathbf{b} \in {\mathcal{A}}^n} {P}(\mathbf{b}) =1$. Note that we already computed the value of 
$\sum_{\mathbf{b} \in {\mathcal{A}}^n} \widetilde{P}(\mathbf{b}) \log \widetilde{P}(\mathbf{b})$ (cf. eq. \ref{pblogpb}).
Also note that nothing would change in the derivation of eq. (\ref{pblogpb}) if we were computing
$\sum_{\mathbf{b} \in {\mathcal{A}}^n} P(\mathbf{b}) \log \widetilde{P}(\mathbf{b})$ instead, meaning that
\begin{equation}
 \sum_{\mathbf{b} \in {\mathcal{A}}^n} \widetilde{P}(\mathbf{b}) \log \widetilde{P}(\mathbf{b})=
\sum_{\mathbf{b} \in {\mathcal{A}}^n} {P}(\mathbf{b}) \log \widetilde{P}(\mathbf{b}).
\end{equation}
We therefore obtain
\begin{equation}
 H_n(\mu)-H_n(\widetilde{\mu}^{(k)}) \leq 0.
\end{equation}
Dividing this by $n$ and taking the limit $n \to \infty$ we obtain inequality (\ref{entropyineq}). $\square$

Let $\mu, \mu_n \in \mathfrak{M}(X)$. If $\int_X f d\mu_n \to \int_X f d\mu$ as
$n \to \infty$ for every bounded, continuous real function $f$ on $X$, we say that
$\mu_n$ \emph{converges weakly} to $\mu$ and write $\mu_n \Rightarrow \mu$. Proof of the following 
useful criterion of weak convergence, originally due to
Kolmogorov and Prohorov \cite{Kolomogorov54}, 
 can be found in \cite{Bilingsley68}.
\begin{thm}\label{weakconvcrit}
 Let $U$ be a subclass of the smallest $\sigma$-algebra containing all open sets of
$X$ such that \emph{(i)} $U$ is closed under the formation of finite intersections and
\emph{(ii)} each open set in $X$ is a finite or countable union of elements of $U$. 
If $\mu_n(A) \to \mu(A)$ for every $A \in U$, then  $\mu_n \Rightarrow \mu$.
\end{thm}
The subclass $U$ satisfying hypothesis of the above theorem is called \emph{convergence determining class}.
It is easy to verify that $\mathit{Cyl}(X)$ is a convergence determining class for measures in $\mathfrak{M}(X)$,
hence the following proposition.
 \begin{prop}
  The sequence of $k$-th order Bayesian approximations of
$\mu \in \mathfrak{M}_\sigma(X)$ weakly converges to  $\mu$ as $k \to \infty$.
 \end{prop}
\emph{Proof:}
Let $n>0$, $\mathbf{b} \in {\mathcal{A}}^n$ and let
$\widetilde{P}_k(\mathbf{b})=\widetilde{\mu}^{(k)}([\mathbf{b}]_0)$, 
 $P(\mathbf{b})=\mu([\mathbf{b}]_0)$.
Since for $k \geq n$ $\widetilde{P}_k(\mathbf{b})=P(\mathbf{b})$,
we obviously have $\lim_{k \to \infty} \widetilde{P}_k(\mathbf{b})=P(\mathbf{b})$.
Theorem \ref{weakconvcrit}, coupled with the fact that $\mathit{Cyl}(X)$ is a convergence determining class
leads to the conclusion that
$\widetilde{\mu}^{(k)} \Rightarrow \mu$.  $\square$ 
\section{Cellular automata}

Let $w: \mathcal{A} \times \mathcal{A}^{2r+1} \to [0,1]$, whose values are denoted by $w(a|\mathbf{b})$
for $a \in \mathcal{A}$, $\mathbf{b} \in \mathcal{A}^{2r+1}$, satisfying
$\sum_{a \in \mathcal{A}} w(a|\mathbf{b})=1$, be called \emph{local transition function}
of \emph{radius} $r$, and its values will be called \emph{local transition probabilities}.
\emph{Probabilistic cellular automaton}  with local 
transition function $w$ is a map $F: \mathfrak{M}(X) \to \mathfrak{M}(X)$ defined as
\begin{equation} \label{rulefed}
(F\mu)([\mathbf{a}]_i)=\sum_{\mathbf{b}\in \mathcal{A}^{|\mathbf{a}|+2r}} w(\mathbf{a}| \mathbf{b}) \mu([\mathbf{b}]_{i-r})
\mathrm{\,\, for\,\, all\,\,}  i \in \mathbb{Z}, \mathbf{a} \in \mathcal{A}^{\star},
\end{equation}
where we define
\begin{equation} \label{defw}
 w(\mathbf{a}| \mathbf{b}) = \prod_{j=1}^{|\mathbf{a}|} w(a_j|b_{j}b_{j+1}\ldots b_{j+2r}).
\end{equation}
When the function $w$ takes values in the set $\{0,1\}$, the corresponding cellular automaton is called 
\emph{deterministic CA}. 

For any probabilistic measure $\mu \in \mathfrak{M}(X)$, we define the orbit of $\mu$ under $F$ as
\begin{equation}
 \{  F^n \mu \}_{n=0}^{\infty}.
\end{equation}
In general, it is very difficult to compute $F^n \mu$ directly, and no general method for doing this is known.
To see the source of the difficulty, let us take ${\mathcal{A}}=\{0,1\}$ and let us consider the example of rule 14, 
for which local transitions probabilities are given by
\begin{align}
w(1|000) = 0, \,  w(1|001) = 1, \,  w(1|010) = 1, \,  w(1|011) = 1,  \nonumber \\
w(1|100) = 0, \,  w(1|101) = 0, \,  w(1|110) = 0, \,  w(1|111) = 0.
\end{align}
Let us further suppose that we want to compute orbit of 
 a shift-invariant Bernoulli measure $\mu_{1/2}$, such that
for any block $\mathbf{b} \in {\mathcal{A}}^{\star}$,
$\mu_{1/2}([\mathbf{b}])=\left(1/2 \right) ^{|\mathbf{b}|}$. If we, for example, consider blocks
$\mathbf{b}$ of length 2, then, defining 
$P_n(\mathbf{b})=(F^n \mu_{\rho})([\mathbf{b}]),$
equation (\ref{rulefed}) becomes 
\begin{align} \label{r14exact}
 P_{n+1}(00)&=
P_n(0000)+
P_n(1000)+
P_n(1100)+
P_n(1101)+
P_n(1110)\nonumber \\
+&P_n(1111), \nonumber \\
P_{n+1}(01)&=
P_n(0001)+
P_n(1001)+
P_n(1010)+
P_n(1011),\nonumber \\
P_{n+1}(10)&=
P_n(0100)+
P_n(0101)+
P_n(0110)+
P_n(0111),\nonumber \\
P_{n+1}(11)&=
P_n(0010)+
P_n(0011).
\end{align}
It is obvious that this system of equations cannot be iterated over $n$, because on the left hand side
we have probabilities of blocks of length 2, and on the right hand side -- probabilities of blocks of
length 4. Of course, not all these probabilities are independent, thus it will be better to 
rewrite the above using short form representation. Since among block probabilities of length 2
only 2 are independent, we can take only two of the above equations, and express all block
probabilities occurring in them by their short form  representation, using eq. (\ref{shortform3}).
This reduces eq. (\ref{r14exact}) to
\begin{align} \label{r14exactreduced}
 P_{n+1}(0)&=1-P_n(0)+P_n(000), \nonumber \\
 P_{n+1}(00)&= 1-2 P_n(0)+P_n(00) + P_n(000).
\end{align}
Although much simpler, the above system of equations still cannot be iterated, because on the right hand
side we have an extra variable $P_n(000)$. To put it differently, one cannot reduce iterations
of $F$  to iterations of a finite-dimensional map (in this case, two-dimensional map).

Before we continue, let us remark that although the aforementioned reduction is not, in general, possible,
one can, nevertheless, in some circumstances compute  $(F^n \mu)([\mathbf{b}])$ for some selected (typically short) blocks
 $\mathbf{b}$ and for some reasonably simple $\mu$. Such calculations use entirely different approach, and typically they exploit features of
 a particular CA rule, thus they  cannot be easily generalized.  
For example, when $\mu$ is a Bernoulli measure, probabilities of
blocks of length up to 3 have been computed for a number of binary cellular automata rules, using the method of preimage
counting \cite{paper34,paper44,paper40,paper39,paper11}. We will, however, not be concerned with these methods here. 
Instead, we will now turn our attention to approximate
methods for computing $F^n \mu$.


Since the reduction of iterations of $F$ to iterations of finitely-dimensional map is, in general,
impossible, we can try to perform this task in an approximate fashion. In the case of rule 14 discussed above, we
could use use Bayesian approximation for this purpose, 
\begin{equation}
 P_n(000) \approx \mfrac{P_n(00) P_n(00)}{P_n(0)}.
\end{equation}
Equations (\ref{r14exactreduced})  would then become
\begin{align}
 P_{n+1}(0)&=1-P_n(0)+\mfrac{P_n(00)^2}{P_n(0)}, \nonumber \\
 P_{n+1}(00)&= 1-2 P_n(0)+P_n(00) +\mfrac{P_n(00)^2}{P_n(0)}.
\end{align}
The above is a formula for recursive iteration of a two-dimensional map, thus one could compute
$P_{n}(0)$ and $P_{n}(00)$ for consecutive $n=1,2 \ldots$ without referring to any other block
probabilities, in stark contrast with eq. (\ref{r14exactreduced}). This, in fact, is the main idea
behind the \emph{local structure approximation} which will be formally introduced in the next section.

\section{Approximate orbits of measures}
Given the difficulty of finding $F^n \mu$, H. Gutowitz \textit{et. al.} \cite{gutowitz87a,gutowitz87b} developed a method 
of approximating orbits of $F$, known as the \emph{local structure theory}.

Following \cite{gutowitz87a}, let us define the \emph{scramble operator} of order $k$, denoted by $\Xi^{(k)}$, to be a map
from $\mathfrak{M}_\sigma (X)$, the set of shift-invariant measures on $X$, to the set of finite block measures
of order $k$, such that
\begin{equation}
\Xi^{(k)} \mu=\widetilde{\mu}^{(k)}.
\end{equation}
The sequence
\begin{equation} \label{approxorbit}
\left\{ \left(\Xi^{(k)} F \Xi^{(k)} \right)^n \mu \right\}_{n=0}^{\infty}
\end{equation}
will be called the \emph{local structure approximation} of level $k$ of the exact orbit $\{  F^n \mu \}_{n=0}^{\infty}$.
Note that all terms of this sequence are Markov measures, thus the entire local structure approximation
of the orbit lies in $\mathfrak{M}^{(k)}(X)$.

The main hypothesis of the local structure theory is that eq. (\ref{approxorbit})
 approximates the actual orbit  $\{  F^n \mu \}_{n=0}^{\infty}$ increasingly well as $k$ increases. The meaning
of ``approximates'' is not rigorously defined in the original paper of H. Gutowitz \textit{et. al.} \cite{gutowitz87a}.
We will shortly prove that every point of the approximate orbit weakly converges to the  corresponding point
of the exact orbit as $k \to \infty$. 
 To do this, we need  the following useful result.
\begin{prop}
Let $k$ be a positive integer and $\mathbf{b} \in {\mathcal{A}}^{\star}$.
If $k\geq |\mathbf{b}|+2r$, then
$$ F \mu ([\mathbf{b}])=F \Xi^{(k)} \mu ([\mathbf{b}])=\Xi^{(k)} F \mu ([\mathbf{b}]).$$
\end{prop}
To prove it, note that $\mu([\mathbf{a}])=\widetilde{\mu}^{(k)}([\mathbf{a}])$ for all blocks 
$\mathbf{a}$ of length up to $k$. The first equality of the proposition can be written as
\begin{equation}
\sum_{\mathbf{a}\in \mathcal{A}^{|\mathbf{b}|+2r}} w(\mathbf{a}| \mathbf{b}) \mu([\mathbf{a}])
=
\sum_{\mathbf{a}\in \mathcal{A}^{|\mathbf{b}|+2r}} w(\mathbf{a}| \mathbf{b}) \widetilde{\mu}^{(k)}([\mathbf{a}]).
\end{equation}
The equality holds when $|\mathbf{a}|\leq k$, that is, $|\mathbf{b}|+2r \leq k$.

The second equality is a result of the fact that the scramble operator only modifies probabilities of blocks of length greater than $k$.
Since $k\geq |\mathbf{b}|+2r$, we have $|\mathbf{b}|<k$ and therefore $F \mu ([\mathbf{b}])=\Xi^{(k)} F \mu ([\mathbf{b}])$.  $\square$ 

Since $F^n$ can be considered as a cellular automaton rule of radius $nr$, 
when $k\geq |\mathbf{b}|+2nr$ we have $F^n \mu ([\mathbf{b}])=F^n \Xi^{(k)} \mu ([\mathbf{b}])$. We can insert
as many $\Xi^{(k)}$ on the right hand side anywhere we want, and nothing will change, because $\Xi^{(k)}$ does
not modify relevant block probabilities. This yields an immediate corollary.
\begin{cor}
Let $k$ and $n$ be  positive integers and $\mathbf{b} \in {\mathcal{A}}^{\star}$.
If $k\geq |\mathbf{b}|+2nr$, then
$$ F^n \mu ([\mathbf{b}])=\left( \Xi^{(k)} F \Xi^{(k)} \right)^n \mu ([\mathbf{b}]).$$
\end{cor}
This means that for a given $n$, measures of cylinder sets in the approximate 
measure $\left( \Xi^{(k)} F \Xi^{(k)} \right)^n \mu$
converge to measures of cylinder sets in $F^n \mu$. By the virtue of Theorem \ref{weakconvcrit} we thus obtain
the following result.
\begin{thm} \label{lstconveergence}
 Let $F$ be a cellular automaton and $\mu$ be a shift-invariant measure in $\mathfrak{M}_{\sigma}(X)$. Let $\nu_n^{(k)}$ be a
local structure approximation of level $k$ of the measure $F^n \mu$, i.e., $\nu_n^{(k)}=\left( \Xi^{(k)} F \Xi^{(k)} \right)^n \mu$.  Then 
for any positive integer $n$, 
$\nu_n^{(k)} \Rightarrow F^n \mu$ as $k \to \infty$.
\end{thm}

\section{Local structure maps}
A nice feature of Markov maps is that they can be entirely described by specifying probabilities
of a finite number of blocks. This makes construction of finite-dimensional maps
generating approximate orbits possible.

If  $\nu_n^{(k)}=\left( \Xi^{(k)} F \Xi^{(k)} \right)^n \mu$, then $\nu_n^{(k)}$ satisfies recurrence equation
\begin{equation}
  \nu_{n+1}^{(k)}= \Xi^{(k)} F \Xi^{(k)} \nu_{n}^{(k)}.
\end{equation}
On both sides of this equation we have measures  in $\mathfrak{M}^{(k)}(X)$, and these are completely determined
by probabilities of blocks of length $k$. If $|\mathbf{b}|=k$, we obtain
\begin{equation}
  \nu_{n+1}^{(k)}([\mathbf{b}])= \Xi^{(k)} F \Xi^{(k)} \nu_{n}^{(k)}([\mathbf{b}]),
\end{equation}
and, since $\Xi^{(k)}$ does not modify probabilities of blocks of length $k$, this simplifies to 
\begin{equation}
  \nu_{n+1}^{(k)}([\mathbf{b}])= F \Xi^{(k)} \nu_{n}^{(k)}([\mathbf{b}]).
\end{equation}
By the definition of $F$,
\begin{equation}
  \nu_{n+1}^{(k)}([\mathbf{b}])= 
\sum_{\mathbf{a}\in \mathcal{A}^{|\mathbf{b}|+2r}} w(\mathbf{a}| \mathbf{b})  
\left( \Xi^{(k)} \nu_{n}^{(k)} \right)
([\mathbf{a}]),
\end{equation}
and, by the definition of Bayesian approximation,
\begin{equation}
  \nu_{n+1}^{(k)}([\mathbf{b}])= 
\sum_{\mathbf{a}\in \mathcal{A}^{|\mathbf{b}|+2r}} w(\mathbf{a}| \mathbf{b})  
\mfrac
{ 
\prod_{i=1}^{2r+1} \nu_{n}^{(k)} ( [\mathbf{a}_{[ i, i+k-1]}])  
}
{ 
\prod_{i=1}^{2r} \nu_{n}^{(k)} ( [\mathbf{a}_{[i+1,i+k-1]}]) 
}.
\end{equation}
To simplify the notation, let us define $Q_n(\mathbf{c})= \nu_{n}^{(k)}([\mathbf{c}])$. Then, 
using consistency conditions in order to obtain on the right hand side expression involving only probabilities of blocks
of length $k$, we rewrite the previous equation to take the form
\begin{equation} \label{lstmapcomponents}
  Q_{n+1}(\mathbf{b})= 
\sum_{\mathbf{a}\in \mathcal{A}^{|\mathbf{b}|+2r}} w(\mathbf{a}| \mathbf{b})  
\mfrac
{ 
\prod_{i=1}^{2r+1} Q_n ( \mathbf{a}_{[ i, i+k-1]})  
}
{ 
\prod_{i=1}^{2r} \sum_{c \in \mathcal{A}} Q_n (c \mathbf{a}_{[i+1,i+k-1]}) 
}.
\end{equation}
The above equation can be written separately for all $\mathbf{b} \in \mathcal{A}^k$.
If we arrange $Q_n(\mathbf{b})$  for all $\mathbf{b} \in \mathcal{A}^k$ in lexicographical order
  to form a vector $\mathbf{Q}_n$, we will obtain
\begin{equation}
 \mathbf{Q}_{n+1} = L^{(k)} \left(\mathbf{Q}_{n}\right),
\end{equation} 
where $L^{(k)}: [0,1]^{|\mathcal{A}|^k} \to [0,1]^{|\mathcal{A}|^k}$ has components defined by eq. (\ref{lstmapcomponents}).
$L^{(k)}$  will be called
 \emph{local structure map} of level $k$. First $N^k-N^{k-1}$ components of $L^{(k)}$
will be denoted by $L^{(k)}_{Top}$, and the remaining components by $L^{(k)}_{Bot}$,
and therefore the local structure map can be written as
\begin{equation}
 \left[ \begin {array}{c} 
\mathbf{Q}^{(k)}_{Top}\\
\mathbf{Q}^{(k)}_{Bot}
\end {array} \right] 
\longmapsto
 \left[ \begin {array}{c} 
L^{(k)}_{Top}  \left(\mathbf{Q}^{(k)}_{Top},\mathbf{Q}^{(k)}_{Bot}\right)\\
L^{(k)}_{Bot} \left(\mathbf{Q}^{(k)}_{Top},\mathbf{Q}^{(k)}_{Bot}\right)
\end{array} \right]. 
\end{equation}
By Proposition \ref{hilowdependenceprop}, $\mathbf{Q}^{(k)}_{Bot}$ can be expressed in terms of
$\mathbf{Q}^{(k)}_{Top}$, 
\begin{equation}
 \mathbf{Q}^{(k)}_{Bot}=
[0,\ldots,0,1]^T
 - \left(\mathbf{B}^{(k)}\right)^{-1} \mathbf{A}^{(k)}\mathbf{Q}^{(k)}_{Top},
\end{equation}
and, therefore, only the ``top'' component of our map needs to be considered,
\begin{equation} \label{LSTlongform}
 \mathbf{Q}^{(k)}_{Top} \longmapsto L^{(k)}_{Top} 
\left(\mathbf{Q}^{(k)}_{Top},
[0,\ldots,0,1]^T
 - \left(\mathbf{B}^{(k)}\right)^{-1} \mathbf{A}^{(k)}\mathbf{Q}^{(k)}_{Top}
\right).
\end{equation}
We will call the above map the \emph{reduced long form} of the local structure map, and write it as
\begin{equation} 
 \mathbf{Q}^{(k)}_{Top} \longmapsto L^{(k)}_{red.\,long} 
\left(\mathbf{Q}^{(k)}_{Top}
\right).
\end{equation}

Now using eq. (\ref{shorttolongtrafo}), we have  $\mathbf{Q}^{(k)}_{Top}=\mathbf{M}^{(k)}_{short} \mathbf{Q}^{(k)}_{short}$,
and  we  can change variables in eq. (\ref{LSTlongform}) from long to short block representation. This yields
\begin{equation} \label{LSTshortform}
\mathbf{Q}^{(k)}_{short} \longmapsto \left(\mathbf{M}^{(k)}_{short} \right)^{-1} L^{(k)}_{red.\,long} 
\left(\mathbf{M}^{(k)}_{short} \mathbf{Q}^{(k)}_{Top}
\right).
\end{equation}
We will call the above map the \emph{reduced short} form of local structure map, and write is as
\begin{equation} 
 \mathbf{Q}^{(k)}_{short} \longmapsto L^{(k)}_{red.\,short} 
\left(\mathbf{Q}^{(k)}_{short}
\right).
\end{equation}

As an example, consider rule 184 given by
\begin{align} \label{wfor184}
w(1|000)=0, \, w(1|001)=0, \, w(1|010)=0, \, w(1|011)=1, \nonumber \\
w(1|100)=1, \, w(1|101)=1, \, w(1|110)=0, \, w(1|111)=1,
\end{align}
and suppose we wish to construct local structure map of level 2 for this rule.
Let $P_n(\mathbf{b})=F^n \mu([\mathbf{b}])$. Using eq. (\ref{rulefed}) we obtain for $r=1$, $|b|=3$
\begin{equation}
 P_{n+1}(\mathbf{b})=\sum_{\mathbf{a}\in \mathcal{A}^{5}} 
 w(\mathbf{a}| \mathbf{b}) P_{n}(\mathbf{a}).
\end{equation}
Using definition of $w(\mathbf{a}| \mathbf{b})$ given in eq. (\ref{defw})
and transition probabilities given in eq. (\ref{wfor184}) we obtain
\begin{align} \label{r184exact}
 P_{n+1}(00)&=
P_n(0000)+
P_n(0001)+
P_n(0010),\nonumber \\
P_{n+1}(01)&=
P_n(0011)+
P_n(0100)+
P_n(0101)+
P_n(1100)+
P_n(1101),\nonumber \\
P_{n+1}(10)&=
P_n(0110)+
P_n(1000)+
P_n(1001)+
P_n(1010)+
P_n(1110),\nonumber \\
P_{n+1}(11)&=
P_n(0111)+
P_n(1011)+
P_n(1111).
\end{align}
This set of equations describes exact relationship between block probabilities at step $n+1$
and block probabilities at step $n$. Note that $3$-block probabilities at step $n+1$ are given in terms of $5$-blocks
probabilities at step $n$, thus it is not possible to iterate these equations.
Local structure map of order $3$, given by eq. (\ref{lstmapcomponents}), becomes
\begin{multline*}
 Q_{n+1}(00)={\mfrac {{Q_{n}(00)}^{3}}{ \left( Q_{n}(00) +Q_{n}(01) \right)^{2}}}+
{\mfrac {{Q_{n}(00)}^{2}Q_{n}(01)}{ \left( Q_{n}(00)+Q_{n}(01) \right)
^{2}}} \\
+{\mfrac {Q_{n}(00)Q_{n}(01)Q_{n}(10)}{ \left( Q_{n}(00)+Q_{n}(01) \right)
\left( Q_{n}(10)+Q_{n}(11) \right) }}, 
\end{multline*}
\begin{multline*}
 Q_{n+1}(01)={\mfrac {Q_{n}(00)Q_{n}(01)Q_{n}(11)}{ \left( Q_{n}(00)+Q_{n}(01) \right)  \left(
Q_{n}(10)+Q_{n}(11) \right) }}  \\
+{\mfrac {Q_{n}(00)Q_{n}(01)Q_{n}(10)}{ \left(
Q_{n}(00)+Q_{n}(01) \right)  \left( Q_{n}(10)+Q_{n}(11) \right) }}   \\
+{\mfrac {{Q_{n}(01)}^{2}Q_{n}(10)}{ \left( Q_{n}(00)+Q_{n}(01) \right) \left(
Q_{n}(10)+Q_{n}(11) \right) }} \\
+{\mfrac {Q_{n}(11)Q_{n}(10)Q_{n}(00)}{ \left(
Q_{n}(00)+Q_{n}(01) \right) \left( Q_{n}(10)+Q_{n}(11) \right) }}  \\
+{\mfrac
{Q_{n}(01)Q_{n}(11)Q_{n}(10)}{ \left( Q_{n}(00)+Q_{n}(01) \right) \left(
Q_{n}(10)+Q_{n}(11) \right) }}, 
\end{multline*}
\begin{multline*}
 Q_{n+1}(10)={\mfrac {Q_{n}(01)Q_{n}(11)Q_{n}(10)}{
\left( Q_{n}(10)+Q_{n}(11) \right) ^{2}}}+{\mfrac {Q_{n}(10){Q_{n}(00)}^{2}}{
\left( Q_{n}(00)+Q_{n}(01) \right) ^{2}}}\\+{\mfrac {Q_{n}(00)Q_{n}(01)Q_{n}(10)}{ \left( Q_{n}(00)+Q_{n}(01) \right) ^{2}}} 
+{\mfrac {
Q_{n}(01){Q_{n}(10)}^{2}}{ \left( Q_{n}(00)+Q_{n}(01) \right)  \left(
Q_{n}(10)+Q_{n}(11) \right) }}\\
+{\mfrac {{Q_{n}(11)}^{2}Q_{n}(10)}{ \left(
Q_{n}(10)+Q_{n}(11) \right)^{2}}},
\end{multline*}
\begin{multline} \label{r184lst2}
 Q_{n+1}(11)={\mfrac {Q_{n}(01){Q_{n}(11)}^{2}}{
\left( Q_{n}(10)+Q_{n}(11) \right) ^{2}}}\\
+{\mfrac {Q_{n}(01)Q_{n}(11)
Q_{n}(10)}{ \left( Q_{n}(00)+Q_{n}(01) \right)  \left( Q_{n}(10)+Q_{n}(11)
\right) }}+{\mfrac {{Q_{n}(11)}^{3}}{ \left( Q_{n}(10)+Q_{n}(11) \right) ^{2}}}.
\end{multline}
Note that eq. (\ref{r184lst2}) can be obtained from eq. (\ref{r184exact}) by replacing 
$P$'s by $Q$'s and expressing every $5$-block probability by its Bayesian approximation of order 3.

According to Corollary \ref{reductioncorollary}, only first two components of $[Q_n(00), Q_n(01)$, $Q_n(10), Q_n(11)]$
are independent, that is, $Q_n(00)$, $Q_n(01)$. This means that we can ignore last two equations in (\ref{r184lst2}),
making in the first two equations substitutions given by eq. (\ref{depbyindep}), that is, 
\begin{align}
 Q_n(10)&=Q_n(01), \\
 Q_n(11)&=1-Q_n(00)-2Q_n(01).
\end{align}
 This yields reduced long form of the local structure map (cf. eq. \ref{LSTlongform}),
\begin{multline*}
 Q_{n+1}(00)={\mfrac {{Q_{n}(00)}^{2}}{Q_{n}(00)+Q_{n}(01)}}+{
\mfrac {Q_{n}(00){Q_{n}(01)}^{2}}{ \left( Q_{n}(00)+Q_{n}(01) \right)
  \left( -Q_{n}(01)+1-Q_{n}(00) \right) }}, 
\end{multline*}
\begin{multline} \label{r184reducedlong}
%
Q_{n+1}(01)=Q_{n}(01)\\
\cdot {\mfrac { \left(2\,{Q_{n}(00)}^{2}-2\,Q_{n}(00)+4\,Q_{n}(00)Q_{n}(01)
+{Q_{n}(01)}^{2}-Q_{n}(01) \right) }{ \left( Q_{n}(00)+Q_{n}(01) \right) 
 \left( Q_{n}(01)-1+Q_{n}(00) \right) }}.
\end{multline}

We now proceed to produce reduced short form of this map. In short block representation, 
we  choose $Q_n(0)$ and $Q_n(00)$ as independent blocks, and probabilities of all other
blocks of length 2 can be expressed by them,
\begin{align*}
Q_n(01) &=Q_n(0)-Q_n(00),\\
Q_n(10) &=Q_n(0)-Q_n(00),\\
Q_n(11) &=Q_n(00)-2 Q_n(0)+1.
 \end{align*}
With this change of variables, eq. (\ref{r184reducedlong}) becomes 
\begin{align} \label{r184redshort}
 Q_{n+1}(0)&= Q_{n}(0), \nonumber \\
 Q_{n+1}(00)&=\mfrac{Q_n(00)^2}{Q_n(0)}
+ \mfrac{Q_n(00)[Q_n(0)-Q_n(00)]^2}{Q_n(0)[1-Q_n(0)]}.
\end{align}
This is the reduced short form of the local structure map (cf. eq. \ref{LSTshortform}). Note that this form
is not only simpler than the original local structure map,
but it also makes it easier to see  an important property of the map, namely the fact that the probability of $0$
is invariant. This actually is true for the orbit of rule 184: probability of 0 (and 1) stays the same
along the orbit. We have here, therefore, an example of a  case where the local structure map ``inherits'' a property of the
rule it approximates. In this case, it inherits the so-called additive invariant. 

Not only does the map inherit the invariant from the exact orbit of rule 184, but it also converges to the
``right'' value. One can easily find its fixed points, determine their stability, and from there determine
$\lim_{n \to \infty} Q_n(00)$.  Since $Q_n(0)$ is constant, let us denote $Q_n(0)=1-\rho$, so that $Q_n(1)=\rho$. 
Eqs. (\ref{r184redshort}) then reduces to
\begin{equation} 
 Q_{n+1}(00)=\mfrac{Q_n(00)^2}{1-\rho}
+ \mfrac{Q_n(00)[Q_n(1-\rho-Q_n(00)]^2}{\rho(1-\rho)}.
\end{equation}
This nonlinear difference equation has three fixed points, $0$, $1-\rho$ and $1-2\rho$. The second one, $1-\rho$,
is always unstable. The first one, $0$, is unstable for $\rho<1/2$, and stable for $\rho>1/2$. The third one,
$1-2 \rho$, is is stable for $\rho<1/2$, and unstable for $\rho>1/2$. We can, therefore, write
\begin{equation}
\lim_{ n \to \infty} Q_n(00)=
\begin{cases}
1-2 \rho, & \rho<1/2 \\
0,  & \rho \geq 1/2.
\end{cases}
\end{equation}
Remarkably, this agrees with the exact limiting value of $P_n(00)=F^n \mu  ([00])$
for rule 184 provided that $\mu$ is a Bernoulli measure, as computed in \cite{paper11}. Again, we can say that the 
local structure map in this case  inherits the limiting value of the probability $P_n(00)$ from the exact orbit.

\section{Conclusions}
We have formalized the idea of local structure theory and demonstrated that orbits of shift-invariant measures
under probabilistic (or deterministic) CA can be approximated by orbits of
 $(N-1)N^k$-dimensional maps, called reduced local structure maps. The paper presented detailed
procedure for construction of such maps. After this foundation has been laid out, further research is
clearly needed. The main question which remains is the relationship between orbits of reduced local structure maps
and exact orbits. Theorem \ref{lstconveergence} reveals one such relationship, namely that points of the 
orbits of the local structure map weakly converge to corresponding points of the exact orbit. Much more, however, can
be said. For example, as we already noticed in the case of rule 184, local structure map ``inherits'' an additive invariant
form the exact orbit. One can prove that this is a general property which holds for arbitrary CA rule with additive 
invariant(s). Are other important  properties of CA, such as, for example, nilpotency or equicontinuity, ``inherited'' 
in a similar fashion? Can we rigorously prove that certain features of exact orbits are preserved when
exact orbits are replaced by local structure approximated orbits? What are these features? These questions are currently under 
investigation and will be  reported in a follow-up paper.

\section{Acknowledgements}
The author acknowledges partial financial support from the Natural
Sciences and Engineering Research Council of Canada (NSERC) in the
form of Discovery Grant. Some calculations on which this work is based were made
 possible by the facilities of the Shared
Hierarchical Academic Research Computing Network (SHARCNET:www.sharcnet.ca) and
Compute/Calcul Canada. 

\providecommand{\href}[2]{#2}\begingroup\raggedright\endgroup
\providecommand{\href}[2]{#2}\begingroup\raggedright\endgroup

\end{document}